\documentclass[secnumarabic,twocolumn,prl,amsfonts]{revtex4}
\usepackage{amsmath}
\usepackage{epsfig}
\usepackage{bm}

\newcommand{\ket}[1]{\ensuremath{|#1\rangle}}
\newcommand{\bra}[1]{\ensuremath{\langle#1|}}
\newcommand{\nuc}[2]{\mbox{${}^{#1}\rm #2$}}
\newcommand{\half}{\mbox{$\frac{1}{2}$}}

\newcommand{\quarter}{\mbox{$\frac{1}{4}$}}
\newcommand{\smhalf}{\raisebox{0.4ex}{$\scriptstyle\frac{1}{2}$}}

\newcommand{\AND}{\textsc{and}}
\newcommand{\OR}{\textsc{or}}
\newcommand{\NOT}{\textsc{not}}
\newcommand{\XOR}{\textsc{xor}}
\newcommand{\NAND}{\textsc{nand}}
\newcommand{\CNOT}{controlled-\NOT}
\newcommand{\CLONE}{\textsc{clone}}
\newcommand{\SWAP}{\textsc{swap}}
\newcommand{\bpi}{\boldsymbol\pi}
\newcommand{\bphi}{\boldsymbol\phi}
\begin{document}
\title{NMR Quantum Computation}
\author{Jonathan A. Jones}
\affiliation{Centre for Quantum Computation, Clarendon Laboratory, Parks Road,
Oxford OX1 3PU, UK}\affiliation{ Oxford Centre for Molecular Sciences, New
Chemistry Laboratory, South Parks Road, Oxford OX1 3QT, UK} \received{31st
July 2000} \maketitle \tableofcontents\newpage
\section{Introduction}
Quantum computation \cite{Bennett:2000, Macchiavello:2000, Bouwmeester:2000}
offers the prospect of revolutionising many areas of science by allowing us to
solve problems well beyond the power of our current classical computers
\cite{Shor:1999}.  In particular a quantum computer would be superb for
simulating the behaviour of other complex quantum mechanical systems
\cite{Feynman:1982, Lloyd:1996}. Although the theory of quantum computation
has been studied for many years, and many important theoretical results have
been obtained, early attempts to actually build even the smallest quantum
computer proved extremely challenging.

In recent years there has been considerable interest in the use of NMR
techniques \cite{Ernst:1987} to implement quantum computations
\cite{Cory:1996, Cory:1997, Gershenfeld:1997, Jones:1998a, Chuang:1998a,
Jones:2000b, Jones:2000c}. It has proved surprisingly simple to build small
NMR quantum computers, and while such computers are themselves too small for
any practical use, their mere existence has brought great excitement to a
field largely deprived of experimental achievements.

\subsection{Structure and scope}
In this article I will describe how NMR techniques may be used to build simple
quantum information processing devices, such as small quantum computers, and
show how these techniques are related to more conventional NMR experiments.
Many tricks and techniques well known from conventional NMR studies can be
applied to NMR quantum computation, and it is likely that some of these will
be applied in any large scale quantum computer which is eventually built.
Conversely, techniques from quantum computation could have applications within
NMR.

It is impossible to explain how NMR may be used to implement quantum
computations without first explaining what quantum computation is, and what it
could achieve.  This in turn requires a brief discussion of classical
reversible computation \cite{Feynman:1996}. In order to reduce these
introductory sections to a reasonable length many technical points have
inevitably been skipped over. Wherever possible I will use traditional product
operator notation \cite{Sorensen:1983} to describe how NMR quantum computers
are implemented; it will sometimes, however, be necessary to use more abstract
quantum mechanical notations \cite{Goldman:1988} to describe what these
computers seek to achieve.

Before the advent of NMR quantum computation, almost all research in this
field was performed within a small community of physicists and mathematicians.
Some important results have never been published as conventional papers, but
simply circulate as electronic preprints; copies of most of these can be
obtained from the quant-ph archive on the LANL e-print server \cite{LANL}.
Similarly, many other papers appear as LANL e-prints long before more formal
publication.

\section{Limits to computation}
Over the last forty years there has been astonishing progress in the power of
computational devices using silicon based integrated circuits.  This progress
is summarised in a set of ``laws'', usually ascribed to Moore, although some
were developed by other people. Over this period the power of computational
devices has roughly doubled every eighteen months, or increased ten fold every
five years. Unfortunately, this extraordinary technological progress may not
continue for much longer, as this increase in computing power requires a
corresponding decrease in the size of the transistors on the chip; this
shrinking process cannot be continued indefinitely as the transistors will
eventually be reduced to the atomic scale.  At current rates of progress it is
estimated \cite{Schulz:1999} that the ultimate limits of this approach will be
reached by about 2012, and any further progress in the power of our computers
will require a radically different approach.  One possible approach is to use
the power offered by quantum computation.

\subsection{Computational complexity}
Even if the problems described above are side\-stepped in some way, there are
still strong limits to the problems which we can solve with current computers.
These limits are derived from the underlying theory of computation itself, and
in particular from computational complexity theory \cite{Welsh:1988}.
Complexity theory considers the classification of mathematical problems
according to how difficult they are to deal with, and seeks to divide them
into those which are relatively easy (tractable) and those which are
uncomfortably difficult (intractable).  Note that complexity theory is not
concerned with problems which we do not know how to solve, or which we know
cannot be solved, but only with problems for which an algorithmic solution
(tractable or intractable) is known.

The classical theory of computation has remained largely unchanged for
decades, with a central role played by the Church--Turing thesis. This asserts
that all physically reasonable models of computation are ultimately equivalent
to one another, so that it is not really necessary to consider any particular
computer when assessing the complexity of a problem; any reasonable model will
do.  In particular, the tractability or otherwise of a problem is independent
of the computational device used.  As we will see, quantum computation
challenges this thesis, as quantum computers appear to be fundamentally more
powerful than their classical equivalents.

To make further progress it is necessary to use a more precise measure of the
complexity of an algorithm.  The usual approach is to determine the
computational resources (most commonly defined as the number of elementary
computational operations) required to implement it. Clearly this measure will
depend on the exact nature of the computational resources available to our
computer, and so this is not directly useful.  A better approach is to
consider not a single isolated problem, but a family of closely related
problems, and to determine how the resources required scale within this
family.  To take a simple example, adding two $n$ digit numbers with pencil
and paper requires $n$ separate additions, together with $n$ carries, and so
the time required for addition scales linearly with $n$. Similarly,
multiplying two $n$ digit numbers by long multiplication requires $n^2$
multiplications and a similar number of additions.

Mathematically, adding two $n$ digit numbers is said to be $O(n)$ (that is, of
order $n$). The exact number of steps required will depend on exactly how
elementary computational steps are counted, but the total number of steps will
always scale linearly with $n$. Similarly, long multiplication can always be
performed in a number of steps which varies quadratically with $n$, and so is
$O(n^2)$.

The complexity of a problem, rather than an algorithm, may be defined as the
complexity of the best known algorithm for solving the problem; thus addition
is $O(n)$.  It might seem that multiplication is $O(n^2)$, but in fact long
multiplication is \emph{not} the best known algorithm; a better approach is
known with complexity about $O(n\log n)$ \cite{Welsh:1988}. Strictly speaking
one should also distinguish between an upper bound on the number of steps
known to be \emph{sufficient} to solve a problem (indicated by $O$), and a
lower bound on the number of steps known to be \emph{required} to solve a
problem (indicated by $\Omega$); a problem, such as addition of $n$ digit
numbers, which is both $O(n)$ and $\Omega(n)$ is denoted as $\Theta(n)$.

Problems, such as addition and multiplication, whose complexity is at worst a
polynomial function of $n$, are said to be \emph{easy}, while problems whose
complexity is worse than a polynomial function of $n$ are said to be
\emph{hard}.  For example, consider the problem of finding the prime factors
of an $n$ digit composite number.  The obvious way to do this is to simply try
dividing the composite number by every number less than its square root; as
there are approximately $\sqrt{10^n}=10^{n/2}$ such trial divisors, this
algorithm has complexity $O(10^{n/2})$.  Better algorithms for factoring are
known, but they all have the same property of \emph{exponential} complexity.

For problems with polynomial complexity, especially those whose complexity is
described by a low power of $n$, such as $n$ or $n^2$, the effort required to
solve a problem increases only slowly with $n$.  Thus it should be possible to
tackle such problems for a reasonable range of input sizes, and a modest
increase in computer power should give a significant increase in this range.
Exponential functions, however, rise extremely rapidly with $n$, and so it
will only be possible to solve problems with exponential complexity for
relatively small input sizes; furthermore a modest increase in computer power
will result in only a tiny increase in the range of problems that can be
solved.

The apparent exponential complexity of factoring is a matter of some
importance, as it underlies many cryptographic schemes in use today, notably
those based on the Rivest--Shamir--Adleman (RSA) approach to public key
cryptography, such as PGP (Pretty Good Privacy) \cite{Schneier:1996}.  These
schemes involve a public key, which anyone may use to encrypt a message, and a
private key, which is required to decrypt such messages. The relationship
between these two keys is equivalent to the relationship between the product
of two large prime numbers and the two prime numbers themselves.  The security
of the system depends on the difficulty of determining the private key from a
long public key, which itself depends on the complexity of factoring.  By
contrast, the computational complexity of encrypting and decrypting messages
is only a polynomial function of the size of the keys.  Thus a small increase
in the amount of effort required to use the cryptographic scheme results in an
enormous increase in its security.

\subsection{Quantum complexity}
Quantum computation offers the possibility of bypassing some of the limits
apparently imposed by complexity theory.  This is because a quantum computer
could implement entirely new classes of algorithms which would allow currently
intractable problems to be solved with ease.

The first serious discussion of quantum computation was by Feynman, who
analysed the difficulty of simulating quantum mechanical systems using
classical computers \cite{Feynman:1982}. This difficulty is well known and
easy to understand; it arises from the enormous freedom available to quantum
systems.  For example, a system of $n$ coupled spin-\half\ nuclei inhabits a
Hilbert space of size $2^n$, and so must be described by a vector with $2^n$
components. Thus it appears that any classical algorithm to simulate the
behaviour of $n$ spin-\half\ particles must have complexity at least $O(2^n)$;
within NMR this corresponds to the well known computational difficulty of
simulating a large strongly coupled spin system. This apparently inevitable
exponential complexity makes the simulation of quantum mechanical systems an
intractable problem.

Despite this apparent complexity, however, coupled spin systems evolve in the
``correct'' manner.  Thus, in some sense, such a spin system appears to have
the capacity to solve a problem which is intractable by conventional classical
means.  Clearly using a system to simulate itself is not a huge step forward,
but Feynman suggested that it might also be possible to use one quantum
mechanical system to simulate another quite different system. Thus an easily
controllable system might be used to simulate the behaviour of another less
well behaved system, while a carefully chosen system might be usable as a
general purpose quantum simulator \cite{Feynman:1982, Lloyd:1996}.

Feynman's ideas appear to have been limited to the simulation of physical
systems, and the ideas he described have more in common with analogue
computers than with current digital computers.  In 1985, however, Deutsch
extended these ideas and described a quantum mechanical Turing machine, which
could act as a general purpose computer \cite{Deutsch:1985}.  Deutsch also
described a (somewhat contrived) problem \cite{Deutsch:1986, Deutsch:1992,
Cleve:1998} which could be solved more efficiently on a quantum mechanical
computer than on any classical computer, suggesting that it might be possible
to use a quantum computer to solve otherwise intractable problems.

Since that time several quantum algorithms \cite{Ekert:1996} have been
developed, the most notable of which is Shor's quantum factoring algorithm
\cite{Shor:1999}; this allows a composite number to be factored with a
complexity only slightly greater than $O(n^2)$, thus rendering factoring
tractable.  As well as being of great mathematical interest, this algorithm
has obvious practical significance, as it poses a threat to many current
cryptographic schemes. The discovery of Shor's algorithm triggered an enormous
increase in research directed at quantum computation and related areas of
quantum information theory.

\section{Atomic computation}
Before describing how quantum mechanical computers can be built, it is useful
to consider how an atomic scale system, such as a group of coupled nuclei,
could be used to implement classical computations \cite{Feynman:1996,
Bennett:1973, Fredkin:1982}. Discussions of this predate suggestions that
quantum devices might have fundamentally different properties.

The basic approach is very simple. Classical computation \cite{Feynman:1996}
is implemented with two state devices, usually called bits, and these two
states can be mapped to the two eigenstates of a quantum mechanical two level
system. The calculation then proceeds by manipulating the states of various
bits such that the final state of some group of bits (the ``output register'')
depends on the result of the desired computation.  As will be shown later,
quantum computation is very similar, except that the bits are not confined to
their eigenstates; this effectively allows many different calculations to be
carried out in parallel.

\subsection{Computational circuits}
\begin{figure}
\epsfig{file=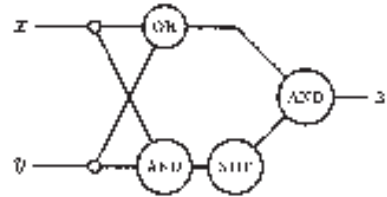} \caption{A circuit to compute the exclusive-\OR\ (\XOR)
function, $z=x\; \text{\XOR}\; y$, using \AND, \OR\ and \NOT\ gates. Note that
this circuit uses three implicit gates, two \CLONE\ gates, shown as small
circles, where wires split into two (to copy the input) and one \SWAP\ gate
(where the wires cross over).  These implicit gates are fairly easy to
implement in traditional electronic computers, but can cause problems in other
designs and so cannot simply be ignored.  Some authors even consider the wires
which interconnect gates as non-trivial gates in their own right.}
\label{fig:XOR}
\end{figure}
Although several different theoretical models are useful for abstract
descriptions of computers, one of the most convenient approaches for
describing how to build small computers is the circuit model, in which bits
interact through gates which implement Boolean logic operations.  One
traditional set of classical gates is the one bit \NOT\ gate together with two
different two bit gates, \AND\ and \OR.  These three gates are said to form an
\emph{adequate} set, in that any desired logic operation can be performed by
building an appropriate circuit (see figure \ref{fig:XOR} for an example)
using some combination of these three gates.

\begin{figure}
\epsfig{file=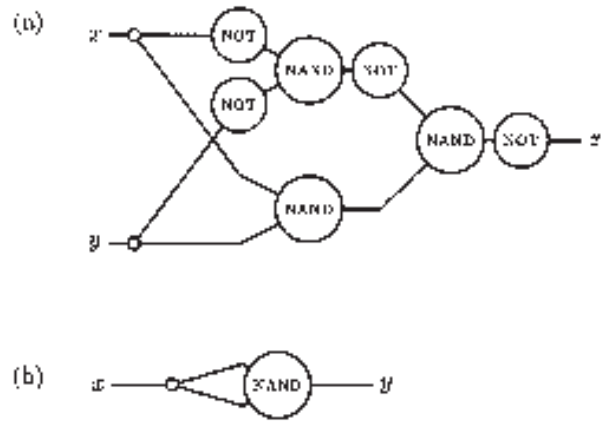} \caption{(a) A circuit to compute the exclusive-\OR\
(\XOR) function, $z=x\; \text{\XOR}\; y$, using only \NAND\ and \NOT\ gates.
This is not the best such circuit; simpler circuits are known, but this one
preserves the basic structure seen in figure \ref{fig:XOR}.  (b) A circuit to
implement a \NOT\ gate, $y=\text{\NOT}\;x$ using a \NAND\ gate. By combining
circuits (a) and (b) it is possible to implement \XOR\ using only \NAND\
gates. Any other function may be computed in a similar fashion, and so \NAND\
is a universal gate. Note, however, that both circuits use implicit \CLONE\
gates, while (a) also uses one implicit \SWAP\ gate.} \label{fig:XOR2}
\end{figure}
In fact it is not necessary to use all these gates: they can themselves all be
obtained using combinations of \NAND\ gates (figure \ref{fig:XOR2}), and so
the \NAND\ gate is \emph{universal} for classical computation. It is, however,
necessary to proceed with some caution, as several other ``implicit'' gates
are also required, such as the \CLONE\ gate, which makes a copy of a bit, and
the \SWAP\ gate, which allows two wires to cross one another.

This description works well with conventional computers, in which bit states
are represented by voltages applied to wires, but it cannot be used with
atomic computers.  Atomic computers represent bit values using the quantum
states of atomic systems, so a logic gate can neither create nor destroy bits;
thus logic gates such as \AND, which have two input bits and only one output
bit, are immediately ruled out.  Similarly, atomic systems evolve under a
series of unitary transformations, which correspond to reversible operations,
while many of the operations described above are clearly not reversible.  In
order to build atomic computers, therefore, it is necessary to use a different
approach, using only reversible logic operations.

\subsection{Reversible computation}
The theory of reversible computation \cite{Feynman:1996, Bennett:1973,
Fredkin:1982} has been studied extensively and, perhaps surprisingly, it is
simple to perform any logic operation in a reversible manner.  The only
irreversible operation required when performing a computation
\cite{Landauer:1961, Landauer:1982} is the preparation of a well defined
initial state, usually taken as having all bits in state 0; after this
initialisation process the computation can be performed entirely reversibly.

The basic approach needed to achieve reversible logic can be summarised in two
simple rules.  First, any logical inputs to a gate must be preserved in the
outputs; this is most simply achieved by copying them to output bits without
change.  Secondly, the output of the gate (here assumed for simplicity to
comprise a single bit) must be combined in a reversible fashion with an
additional auxiliary bit, for example by adding the two bits together using
binary arithmetic modulo two.

\begin{figure}
\begin{center}
\begin{picture}(60,45)
\put(10,40){\line(1,0){40}} \put(30,40){\line(0,-1){35}}
\put(30,40){\circle*{5}} \put(10,10){\line(1,0){40}}
\put(30,10){\circle{10}} \put(0,38){$x$} \put(0,8){$y$}
\put(55,38){$x'$} \put(55,8){$y'$}
\end{picture}
\end{center}
\caption{The \CNOT\ gate, which plays a central role in reversible and quantum
computation.  The target bit is marked by a $\oplus$ symbol, indicating the
close relationship with binary arithmetic modulo two; the control bit is
marked by a dot and a vertical control line (the dot is important in drawings
of larger circuits where a control line may have to cross lines representing
other bits).} \label{fig:CNOT}
\end{figure}
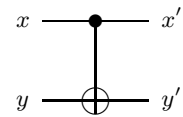

Binary arithmetic modulo two, usually indicated by the symbol $\oplus$, has
the useful properties that $x \oplus 0 = 0 \oplus x = x$ and $x \oplus 1 = 1
\oplus x = \mbox{\sc not}(x)$, while $x \oplus x=0$. A simple example of a
reversible logic gate based on this approach is the \CNOT\ gate (see figure
\ref{fig:CNOT}), which has two inputs, $x$ and $y$ and two corresponding
outputs $x'$ and $y'$. The first input bit is copied to its output bit, so
$x'=x$, and is also combined with the second input bit to give $y'=x\oplus y$.
Thus a \NOT\ gate is applied to the second bit (the target bit) if and only if
the first bit (the control bit) is in state 1. Note that a \CNOT\ gate is
completely reversible; indeed it can be reversed by simply applying the same
gate again (that is, \CNOT\ is its own inverse). Furthermore, as $x\oplus y =
x\; \text{\XOR}\; y$ the \CNOT\ gate is just a reversible \XOR\ gate.

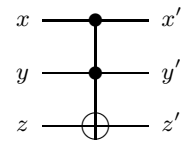
\begin{figure}
\begin{center}
\begin{picture}(60,55)
\put(10,50){\line(1,0){40}} \put(30,50){\line(0,-1){45}}
\put(30,50){\circle*{5}} \put(10,30){\line(1,0){40}}
\put(30,30){\circle*{5}} \put(10,10){\line(1,0){40}}
\put(30,10){\circle{10}}
\put(0,48){$x$}\put(0,28){$y$}\put(0,8){$z$}
\put(55,48){$x'$}\put(55,28){$y'$}\put(55,8){$z'$}
\end{picture}
\end{center}
\caption{The Toffoli gate, which gives $x'=x$, $y'=y$, and
$z'=z\oplus(\text{$x$ {\sc and} $y$})$.} \label{fig:Toffoli}
\end{figure}

A more complex example is provided by the controlled-\CNOT\ gate, often called
the Toffoli gate (figure \ref{fig:Toffoli}), which plays a central role in
reversible logic. This gate has three inputs, $x$, $y$ and $z$, and three
corresponding outputs, $x'$, $y'$ and $z'$. The first two input bits, which
are the logical inputs, are copied unchanged, so that $x'=x$ and $y'=y$. A
\NOT\ gate is then applied to the third bit if and only if \emph{both} $x$ and
$y$ are in state 1; hence $z'=z\oplus(\text{$x$ {\sc and} $y$})$. Thus this
gate can be considered as a reversible equivalent to the conventional \AND\
and \NAND\ gates: to evaluate \mbox{$x$ \sc and $y$} just use a Toffoli gate
with $z=0$, while for a \NAND\ gate set $z=1$.

\begin{figure}
\begin{center}
\raisebox{10ex}{(a)}\quad
\begin{picture}(60,45)
\put(10,40){\line(1,0){40}} \put(30,40){\line(0,-1){35}}
\put(30,40){\circle*{5}} \put(10,10){\line(1,0){40}}
\put(30,10){\circle{10}} \put(0,38){$x$} \put(0,8){$0$}
\put(55,38){$x$} \put(55,8){$x$}
\end{picture}\qquad
\raisebox{10ex}{(b)}\quad
\begin{picture}(100,45)
\put(10,40){\line(1,0){80}}\put(10,10){\line(1,0){80}}
\put(30,40){\circle*{5}}\put(30,40){\line(0,-1){35}}\put(30,10){\circle{10}}
\put(50,10){\circle*{5}}\put(50,45){\line(0,-1){35}}\put(50,40){\circle{10}}
\put(70,40){\circle*{5}}\put(70,40){\line(0,-1){35}}\put(70,10){\circle{10}}
\put(0,38){$x$}\put(0,8){$y$}\put(95,38){$y$}\put(95,8){$x$}
\end{picture}\end{center}
\caption{(a) The \CNOT\ gate can be used to reversibly clone a
bit; (b) three \CNOT\ gates implement a \SWAP\ gate.}
\label{fig:CS}
\end{figure}
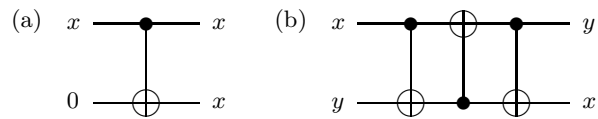

The combination of the Toffoli gate with the \CNOT\ and simple \NOT\ gates
forms an adequate set, in that it is possible to build any other gate using
only a combination of these gates (in particular \CNOT\ gates can also be used
to build the two implicit gates, \CLONE\ and \SWAP, as shown in figure
\ref{fig:CS}). Indeed, the Toffoli gate is universal in its own right, as a
Toffoli gate can be easily converted into a \CNOT\ gate by setting $x=1$, and
to a simple \NOT\ gate by setting $x=y=1$.

\subsection{Reversible function evaluation}
\begin{figure}
\begin{center}
\begin{picture}(65,55)
\put(10,40){\line(1,0){10}} \put(30,30){\line(0,-1){25}}
\put(20,30){{\framebox(20,20){$f$}}} \put(40,40){\line(1,0){10}}
\put(10,10){\line(1,0){40}} \put(30,10){\circle{10}}
\put(0,38){$a$} \put(0,8){$b$} \put(55,38){$a'$} \put(55,8){$b'$}
\end{picture}
\end{center}
\caption{The $f$-\CNOT\ gate, for reversible function evaluation:
$a'=a$ and $b'=b\oplus f(a)$.} \label{fig:fCNOT}
\end{figure}
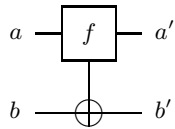

Central to reversible computation is the idea of reversible function
evaluation.  I will initially assume that the function has a single bit as
both input and output, but the generalisation to more complex functions is
straightforward. This can be achieved by constructing a circuit with two
inputs and two outputs, as shown in figure \ref{fig:fCNOT} (an $f$-\CNOT) and
setting $b=0$. The two values of the function, $f(0)$ and $f(1)$, can then be
evaluated by setting $a=0$ and $a=1$ respectively. The $f$-\CNOT\ gate can
itself be built out of simpler gates, such as those described above; in most
cases this will also require a number of \emph{ancilla} bits to hold
intermediate results.  For simplicity these ancilla bits are usually omitted
from diagrams such as figure \ref{fig:fCNOT}.

\section{Quantum computation}
We now have all the basic elements needed to describe how quantum computers
could be used to extend our computational abilities. While large scale quantum
algorithms, such as Shor's algorithm, are too complex to describe here, the
basic ideas are relatively simple.

As described above, a quantum mechanical two-level system can be used to build
a reversible classical computer by using the two eigenstates of the system to
represent bits in logical states 0 and 1.  For example, the two Zeeman levels
of a spin-\half\ nucleus in a magnetic field, \ket{\alpha} and \ket{\beta},
would be suitable for this purpose.  For simplicity the two states are usually
denoted \ket{0} and \ket{1}, allowing quantum computation to be described
without reference to any particular implementation; this choice of basis set
is called the \emph{computational basis}. The system will not be confined to
these two eigenstates, however, but can also be found in superpositions, such
as
\begin{equation}
\frac{\ket{0}+\ket{1}}{\sqrt{2}} \label{eq:sup}
\end{equation}
(the $\sqrt{2}$ term is necessary to ensure that the state is normalised).
For this reason a quantum mechanical two level system has much more freedom
than a classical bit, and so is called a quantum bit, or \emph{qubit} for
short.  A qubit in a superposition state is (in some sense) in both of the
states contributing to the superposition at the same time, so a qubit can
simultaneously occupy two different logical states.

Computational circuits are implemented within quantum computers by performing
physical manipulations so that the computer evolves under a propagator which
implements the desired unitary transformation.  Just as circuits can be built
up from gates these propagators can be assembled from simpler elements, and so
these propagators are often referred to as circuits, even though their
physical implementation may bear little resemblance to conventional electrical
circuits.  As before, this abstract model allows quantum computers to be
described in a device-independent fashion.

As discussed below, it is possible to construct any quantum circuit by
combining a small number of simple propagators, usually called gates.  These
gates only affect one or two qubits at a time, and so it is perfectly
practical to describe their propagators explicitly, for example as a matrix.
A matrix description clearly depends on the choice of basis set, but the usual
choice is to work in the computational basis, in which the basis vectors
correspond with the different logical states of the computer; thus for a two
qubit gate the basis set is $\{\ket{00}, \ket{01}, \ket{10}, \ket{11} \}$. In
many proposed physical implementations of quantum computers, such as NMR, this
basis set is also the natural basis for the system, as the basis states are
eigenstates of the background Hamiltonian.

\subsection{Quantum parallelism}
The central feature underlying all quantum algorithms is the idea of quantum
parallelism, which in turns stems from the ability of quantum systems to be
found in superposition states.  Consider once again the reversible circuit for
function evaluation (figure \ref{fig:fCNOT}).  This can be achieved by
constructing a propagator, $U_f$, applied to two qubits, which performs
\begin{equation}
\ket{a}\ket{b}\stackrel{U_f}\longrightarrow\ket{a}\ket{b\oplus
f(a)}
\end{equation}
and setting $\ket{b}=\ket{0}$.  The two values of the function, $f(0)$ and
$f(1)$, can then be evaluated by setting $\ket{a}=\ket{0}$ and
$\ket{a}=\ket{1}$ as before.  Now consider the effect of applying this circuit
when the first qubit is in a superposition described by equation \ref{eq:sup}
and the second qubit is in state \ket{0} . This can be easily calculated as
quantum mechanics is \emph{linear}, and so the effect of applying a gate to a
superposition is a superposition of the results of applying the gate to the
two eigenstates.  Hence the result is
\begin{equation}
\frac{\left(\ket{0}+\ket{1}\right)}{\sqrt{2}}\ket{0}
\stackrel{U_f}\longrightarrow
\frac{\ket{0}\ket{f(0)}+\ket{1}\ket{f(1)}}{\sqrt{2}}.
\label{eq:fsup}
\end{equation}
Thus the quantum computer has \emph{simultaneously} evaluated the
values of $f(0)$ and $f(1)$.

When applied to more complex systems, quantum parallelism has
potentially enormous power.  Consider a function for which the
input is described by $n$ bits, so that there are $2^n$ possible
inputs (since $n$ bits can be used to describe any integer between
0 and $2^n-1$); a quantum computer using $n$ qubits as inputs can
evaluate the function over all of these $2^n$ inputs in one step.
In effect a quantum computer with $n$ input qubits appears to have
the computational power of $2^n$ classical computers acting in
parallel.  Unfortunately it is not always possible to use this
quantum parallelism in any useful way.  Performing parallel
function evaluation over $n$ qubits will result in a state of the
form
\begin{equation}
\frac{\sum_{i=0}^{2^n-1}\ket{i}\ket{f(i)}}{2^{n/2}},
\end{equation}
which is a superposition of the $2^n$ possible inputs, each \emph{entangled}
with its own function value.  If any attempt is made to measure the state of
the system, then the superposition will collapse into one of its component
values, \ket{r}\ket{f(r)}, where the value of $r$ is chosen at random.  Thus
even though it seems possible to evaluate the function over its $2^n$ input
values, it is only possible to obtain one of these values.

It is, however, sometimes possible to obtain useful information in
a more subtle way.  In some cases, the answer of interest does not
depend on specific values, $f(r)$, but only on global properties
of the function itself.  This is the basis of both Deutsch's toy
algorithm and Shor's quantum factoring algorithm.

\subsection{Deutsch's algorithm}
Deutsch's algorithm \cite{Deutsch:1986, Deutsch:1992, Cleve:1998} was the
first quantum algorithm to be discovered, and is one of the few quantum
algorithms simple enough to describe here.  The problem can be described in
terms of function evaluation, but a more concrete picture can be obtained by
thinking about coins.  Normal coins have two different faces, conventionally
called \emph{heads} and \emph{tails}, but fake coins can be obtained which
have the same pattern on both faces.

Consider an unknown coin, which could be either a normal coin or a fake coin.
In order to determine which type it is, it would seem necessary to look at
both sides to find out whether they showed heads or tails, and then see
whether these two results were the same (a false coin) or different (a true
coin).  With a quantum device, however, it would be possible to look at both
sides simultaneously, and thus determine whether the coin was normal or fake
in a single glance.

The trick lies in abandoning any attempt to determine the pattern shown on
either side of the coin; instead one must simply ask whether the two faces are
the same or different.  This is a property not of the individual faces of the
coin, but of the whole coin, and thus may be extracted from a state of the
kind described by equation~\ref{eq:fsup}. A more detailed explanation of this
approach is given in Section \ref{sec:algs}.

\subsection{Quantum gates}
Just as classical reversible computations can be performed using circuits
built up out of reversible gates, quantum circuits can be constructed using
quantum gates \cite{DiVincenzo:1998b}.  Unlike classical circuits, however,
quantum circuits can include gates which generate and analyse qubits which are
in superpositions of states.

One such gate is the single qubit Hadamard gate, \textsc{H}, which
implements the transformations
\begin{equation}
\ket{0}\stackrel{\textsc{H}}{\longrightarrow}(\ket{0}+\ket{1})/\sqrt{2}
\label{eq:H0}
\end{equation}
\begin{equation}
\ket{1}\stackrel{\textsc{H}}{\longrightarrow}(\ket{0}-\ket{1})/\sqrt{2}
\label{eq:H1}
\end{equation}
As discussed below, this is closely related to a $90^\circ$ pulse,
but differs in some subtle ways; in particular the Hadamard is its
own inverse.

The Hadamard gate is useful as it takes a qubit in an eigenstate to a uniform
superposition of states.  By analogy one can define multi-qubit Hadamard gates
which take a quantum register into a uniform superposition of all its possible
values:
\begin{equation}
\ket{00}\stackrel{\textsc{H}}{\longrightarrow}
(\ket{00}+\ket{01}+\ket{10}+\ket{11})/2.
\end{equation}
This can be easily achieved by applying a one qubit Hadamard to each qubit.

Another important property of the Hadamard gate can be seen by examining the
right hand sides of equations \ref{eq:H0} and \ref{eq:H1}.  These differ only
by the presence of a minus sign, so the two superpositions differ only by the
phase with which the two eigenstates are combined.  The ability of the
Hadamard gate to convert such phase differences into different eigenstates
plays a central role in many quantum algorithms.

\subsection{Universality of quantum gates}
The gates we have seen so far are enough to explain some simple quantum
algorithms; for example, Deutsch's algorithm can be described using only
Hadamard gates and reversible function evaluation gates.  It is, however,
useful to consider other more general quantum gates; indeed as any unitary
transformation can be considered as a quantum gate, we may need to consider an
infinite number of gates.

Within classical models of computation (both reversible and irreversible) it
is possible to construct any desired gate by combining copies of a small
number of simple gates. A similar situation applies in quantum computation,
but in this case there are an infinite number of possible gates (even if we
restrict ourselves to single qubit gates, any rotation around any axis
constitutes a valid one qubit gate). Clearly it cannot be possible to
construct an infinite number of different gates by combining a finite number
of simpler gates, but it \emph{is} possible to simulate any gate to any
desired accuracy \cite{Deutsch:1989, Barenco:1995a}, which is good enough.
Perhaps surprisingly there exists a very large number of two qubit gates which
are \emph{universal} in this restricted sense, in that it is possible to
simulate any desired gate (that is, any unitary transformation) using only one
of these universal gates together with its twin, obtained by swapping the
roles of the two qubits. Indeed, mathematically speaking, almost all two qubit
gates are universal \cite{Barenco:1995c, Deutsch:1995, Lloyd:1995}.

While mathematically interesting, this result is of little immediate practical
implication for most possible implementations of a quantum computer, as it is
usually more sensible to use a larger and more convenient set of gates.  As
one qubit gates are usually much simpler to perform than gates involving two
or more qubits, it is often reasonable to assume that \emph{any} one qubit
gate (or, at least a reasonable approximation to it) is available. The
combination of this set of one qubit gates with any single non-trivial two
qubit gate, such as the controlled-\textsc{not} gate forms an adequate set
\cite{Barenco:1995a}, from which any other gate may be built with relative
ease.

\section{Building NMR quantum computers}\label{sec:building}
While it would in principle be possible to use a wide range of different
approaches to build a quantum computer, all the main proposals to date
\cite{Macchiavello:2000, Bouwmeester:2000} have used broadly similar
approaches, based on the quantum circuit model outlined above.  This model
contains five major components, each of which must be implemented in order to
construct a working computer \cite{DiVincenzo:2000}.  Four central components
can all be implemented within NMR systems as described below, while the fifth
component, error correction, is discussed in Section \ref{sec:phenomena}.

\subsection{Qubits}
The first of these requirements, a set of qubits, appears easy to achieve, as
the two spin states of spin-\half\ nuclei in a magnetic field provide a
natural implementation.  However, one important feature which distinguishes
NMR quantum computers from other suggested implementations is that NMR studies
not a single isolated quantum system, but rather a very large number
(effectively an ensemble) of such systems.  Thus an NMR quantum computer is
actually an ensemble of indistinguishable computers, one on each molecule in
the NMR sample.  This has a number of subtle and important consequences as
discussed below.

\subsection{Logic gates}
In order to perform an arbitrary computation it is necessary to implement
arbitrary quantum logic circuits.  This can be achieved as long as it is
possible to implement an adequate set of gates, which can be combined together
to implement any other desired gate.  While many different sets of gates are
possible, a simple approach is to implement the set of all possible one qubit
gates, together with one or more non-trivial two qubit gates
\cite{Barenco:1995a}.

One qubit gates correspond to rotations of a spin within its own one-spin
Hilbert space, which can be readily achieved using RF fields.  Note that it is
necessary to apply these rotations selectively to individual qubits.  In most
other suggested implementations of quantum computation
\cite{Macchiavello:2000, Bouwmeester:2000} this is easily achieved using some
type of spatial localisation: the physical objects implementing the qubits are
located at well defined and distinct locations in space.  This approach is not
possible in NMR, as each qubit is implemented using an ensemble of nuclei,
each of which is located at a different place in the NMR sample, and all of
which are undergoing rapid motion.  Instead different qubits are implemented
using different nuclei in the same molecule, and they are distinguished using
the different resonance frequencies of each nucleus.

Two qubit gates, such as the controlled-{\sc not} gate, are more complicated
as they involve conditional evolution (that is, the evolution of one spin must
depend on the state of the other spin), and thus require some interaction
between the two qubits. The J-coupling in NMR is well suited to this purpose.
Note that all the different nuclei making up an NMR quantum computer must
participate in a single coupling network. It is not necessary (or even
desirable) that all the nuclei are directly coupled together, but they must be
connected, directly or indirectly, by some chain of resolved couplings.  Since
J-coupling only occurs within a molecule, and does not connect different
molecules, we can treat an ensemble of molecules as an ensemble of identical
mutually isolated computers.

\subsection{Initialisation}
Quantum logic gates transform qubits from one state to another, but this is
only useful if the qubits start off in some well defined initial state.  In
practice it is sufficient to have some method for reaching any one initial
state, and the obvious choice is to have all the qubits in the \ket{0} state,
corresponding to a {\sc clear} operation.  Any other desired starting state
can then be easily obtained.

When, as for NMR, the computational basis coincides with the natural basis of
the quantum system it should in principle be easy to implement {\sc clear} as
it takes the quantum computer to its energetic ground state, and this can be
achieved by some cooling process. Unfortunately this approach is not practical
in NMR as the Zeeman energy gap is small compared with the Boltzman energy at
any reasonable temperature; thus at room temperature the population of all the
states will be almost equal, with only small deviations (around one part in
$10^4$) from the average. Techniques for enhancing spin polarization
\cite{Jones:2000d}, such as optical pumping \cite{Walker:1997, Navon:1996,
Pietrass:1999}, and the use of \emph{para}-hydrogen \cite{Natterer:1997,
Duckett:1999, Hubbler:2000} allow this deviation to be increased, but with the
exception of optically pumped noble gases it has so far proved impossible to
even approach a pure ground state system.

This apparent inability to implement the {\sc clear} operation led to NMR
being rejected as a practical technology for implementing quantum computers.
Recently, however, it was realised \cite{Cory:1996} that this conclusion was
over hasty, as with an ensemble quantum computer it is not actually necessary
to produce a pure ground state; instead it suffices to produce a state which
behaves in the same manner as the pure ground state.  This point can be
clarified by considering the density matrix describing a single isolated
spin-half nucleus. This exhibits nearly equal populations for the two
eigenstates, but with a slight excess in the (low energy) \ket{0} state
compared with the (slightly higher energy) \ket{1} state.  No NMR signal will
be observed from the equal populations, as the signals from different
molecules will cancel out, but a small signal can be seen which arises from
the deviations away from the average. Thus, ignoring questions of signal
intensity, for a single isolated nucleus the thermodynamic equilibrium state
is indistinguishable from a pure \ket{0} state.

States of this kind are often called \emph{pseudo-pure states}, or
\emph{effective pure states} \cite{Cory:1996, Cory:1997, Gershenfeld:1997}.
Unfortunately the simple approach outlined above does not work for larger spin
systems, as the pattern of population deviations is more complicated, and does
not have the desired form. Several different techniques have, however, been
developed to tackle this problem.

\subsection{Readout}
The last stage in any quantum computation is to characterise the final state
of the system, so that the result of the computation may be read out.  Just as
for initialisation, a range of different approaches have been used, but all
these approaches combine two major elements.  For simplicity I will assume
that the computation ends with the result qubits in eigenstates; thus it is
only necessary to determine whether a given qubit is in (the pseudo-pure)
state \ket{0} or \ket{1}.

The simplest approach is to apply a $90^\circ$ pulse to the corresponding
spin, and observe the NMR spectrum \cite{Jones:1998a}.  Since \ket{0}
corresponds to the ground state, a qubit in \ket{0} will give rise to an
absorption line; correspondingly a qubit in state \ket{1} will give an
emissive signal.  It is, of course, necessary to acquire some sort of
reference signal, in order to distinguish between these two extremes, but this
can be easily achieved by acquiring the spectrum of the pseudo-pure initial
state.

The second major approach \cite{Chuang:1998a} is to determine the state of one
qubit by analysing the multiplet structure within the spectrum of a
neighbouring spin. If several spins are coupled together, then individual
lines within a multiplet can be assigned to specific states of these
neighbours.  Thus, the spectrum of one spin can give information on the states
of several different qubits.

\subsection{Some two spin systems}
While a number of different systems have been used to build small NMR quantum
computers, all their major features can be explored using two different
two-qubit systems which were used in the earliest demonstrations of NMR
quantum computation \cite{Jones:1998a, Chuang:1998a}.  The most important
difference between these systems is that one uses a homonuclear two-spin
system, while the other is heteronuclear.

\begin{figure}[tb]
\epsfig{file=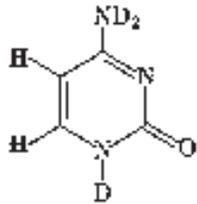} \caption{The structure of partially deuterated cytosine
obtained by dissolving cytosine in $\rm D_2O$; the three protons bound to
nitrogen nuclei exchange with solvent deuterons, leaving two \nuc{1}{H} nuclei
as an isolated two spin system (all other nuclei can be ignored).}
\label{fig:cytosine}
\end{figure}

The first example system uses the two \nuc{1}{H} nuclei of partially
deuterated cytosine in $\rm D_2O$ (see figure \ref{fig:cytosine}). As this
system is homonuclear it is possible to excite both nuclei with a single hard
pulse, and to observe both nuclei in the same spectrum. Another more subtle
advantage is that the pattern of Boltzmann populations is simpler in
homonuclear systems than in their heteronuclear counterparts. There are,
however, two significant disadvantages of such as system.  Firstly the two
\nuc{1}{H} multiplets have relatively similar frequencies, as they lie only
about $\rm1.51\,ppm$ apart, and thus it is necessary to use soft frequency
selective pulses \cite{Freeman:1997} (or sequences of hard pulses and delays
with equivalent effects) in order to address the spins individually. Secondly,
the J-coupling between the two spins is relatively small (about $\rm7\,Hz$),
and so controlled gates take a fairly long time to implement.  It would, of
course, be possible to choose a different molecule, in which the chemical
shift difference or J-coupling was larger, but it is difficult to improve one
without making the other worse. While it is unlikely that cytosine is the
absolutely optimal choice, no other homonuclear \nuc{1}{H} system would be
very much better.

The heteronuclear alternative is probably the most widely used two qubit NMR
system.  It is based on the \nuc{1}{H} and \nuc{13}{C} nuclei in
\nuc{13}{C}-labeled chloroform.  This has the huge advantage that it is
possible to separately excite the two spins using hard pulses, rendering
selective excitation essentially trivial. Furthermore, the relatively large
size of the J-coupling allows two qubit gates to be performed much more
rapidly than in homonuclear systems.  In this heteronuclear system it is not
possible to acquire signals from both spins simultaneously, but this is not a
major problem as it is possible to determine the states of both spins by
examining either the \nuc{1}{H} or the \nuc{13}{C} spectrum.  Similarly, the
complex pattern of populations over the four energy levels of this system does
not fit with the original scheme for generating pseudo-pure states; however,
some more modern schemes are in fact simpler to implement in heteronuclear
systems.

Considering all these issues together, it is not easy to say whether it is
better to use homonuclear or heteronuclear systems to implement two qubit NMR
quantum computers: heteronuclear systems are perhaps simpler to work with, but
homonuclear systems give more elegant results.  With larger spin systems the
issues become even more complex, and a wide range of options have been
explored. It is clear, however, that the simplest approach of using a fully
heteronuclear spin system is unlikely to be practical beyond five qubit
systems, as there are only 5 ``obvious'' spin-half nuclei which can be used
(\nuc{1}{H}, \nuc{13}{C}, \nuc{15}{N}, \nuc{19}{F} and \nuc{31}{P}).  In
practice NMR quantum computers with more than three qubits are likely to
include two or more spins of the same nuclear species; it is, therefore,
essential to consider how computation can be performed in homonuclear systems.

\subsection{Scaling the system up}
The requirements outlined above are adequate for building \emph{small} quantum
computers, suitable for simple demonstrations of quantum information
processing.  If, however, one wishes to build a large scale quantum computer,
suitable for performing interesting computations, then it is necessary to
consider whether the approaches used are limited to such small systems, or
whether (and if so, how) they can be scaled up.  A fifth requirement for
practical quantum computation \cite{DiVincenzo:2000}, the implementation of
fault-tolerant quantum error correction, is described in Section
\ref{sec:phenomena}.

This is an important practical question, but not one which will be addressed
in detail here.  The problems of scaling up NMR quantum computers are
formidable, and have been well described elsewhere \cite{Jones:2000d,
Warren:1997, Gershenfeld:1997b}. Most authors now agree that NMR approaches
are likely to be limited to computers containing 10--20 qubits; this is
significantly smaller than estimates of the size required to perform useful
computations (50--300 qubits).  Furthermore the apparent inability of NMR
systems to perform efficient quantum error correction rules out their use for
many types of problem.

The fundamental difficulties involved in scaling up current NMR quantum
computers to large sizes have led some authors to suggest that this approach
does not actually implement real quantum computation at all.  This is a quite
subtle question which will be discussed further in Section \ref{sec:NMRqm}
below.

\section{Qubits and NMR spin states}
Traditional designs for quantum computers comprise a number of two-level
systems which interact with one another and have some specific interaction
with the outside world, through which they can be monitored and controlled,
but are otherwise isolated.  NMR systems are rather different: a typical NMR
sample comprises not one spin-system, but a very large number of copies, one
from each molecule in the sample, effectively forming an ensemble of copies.
Traditional quantum computers are usually described using Dirac's bra(c)ket
notation \cite{Goldman:1988}, but NMR systems are better described using
density matrices, usually written in the product operator basis
\cite{Sorensen:1983}, which has a number of important consequences. It is
possible to draw close analogies between the states of traditional quantum
computers and those used in descriptions of NMR systems \cite{Jones:1998d},
but it is necessary to proceed with caution.

\subsection{One qubit states}
A single qubit can be in either of its two eigenstates, \ket{0} and \ket{1},
or in some linear superposition of them.  Such a state is most conveniently
written as a column vector in Hilbert space, for example
\begin{equation}
\ket{\psi}=\begin{pmatrix} c_0 \\ c_1 \end{pmatrix}.
\end{equation}
NMR quantum computers cannot be properly described in this way;
instead they must be described using the corresponding density
matrix
\begin{equation}
\rho=\ket{\psi}\bra{\psi}=\begin{pmatrix} c_0^*c_0 & c_1^*c_0 \\
c_0^*c_1 & c_1^*c_1
\end{pmatrix}
\end{equation}
which can then be decomposed as a sum of the four Pauli basis states or their
product operator equivalents, $\half E$, $I_x$, $I_y$, and $I_z$.

Consider first the eigenstates, \ket{0} and \ket{1}, which correspond to the
density matrices
\begin{equation}
\ket{0}\bra{0}=\begin{pmatrix} 1 & 0 \\ 0 & 0 \end{pmatrix}=\half
E+I_z \label{eq:rho0}
\end{equation}
and
\begin{equation}
\ket{1}\bra{1}=\begin{pmatrix} 0 & 0 \\ 0 & 1 \end{pmatrix}=\half
E-I_z \label{eq:rho1}
\end{equation}
respectively.  As all NMR observables are traceless, multiples of the unit
matrix can be added to density matrices at will, and so as far as any NMR
experiment is concerned the density matrix $I_z$ is equivalent to \ket{0},
while $-I_z$ is equivalent to \ket{1}. In the language introduced above, $I_z$
and $-I_z$ are pseudo-pure states, corresponding to \ket{0} and \ket{1}
respectively.  This approach cannot, however, be extended to larger spin
systems without modifications.

Next consider superpositions, such as
$(\ket{0}+\ket{1})/\sqrt{2}$, with its corresponding density
matrix
\begin{equation}
\renewcommand{\arraystretch}{1.2}
\begin{pmatrix} \half & \half \\ \half & \half \end{pmatrix}
=\half E+I_x. \label{eq:rhox}
\end{equation}
As before multiples of the unit matrix can be ignored, and so
$(\ket{0}+\ket{1})/\sqrt{2}$ is equivalent to $I_x$.  Similarly
$\ket{0}+i\ket{1}$ is equivalent to $I_y$, while $\ket{0}-\ket{1}$
is equivalent to $-I_x$. Just as single qubit eigenstates are
closely related to one spin magnetizations, their superpositions
are closely related to one spin coherences.

\subsection{Two qubit states}
While there is a simple relationship between qubit states and NMR states for a
single qubit (a one spin system), this relationship is more complicated in
systems with two or more qubits \cite{Jones:1998d}.  Typically quantum
algorithms start with all qubits in state \ket{0}, which for a two-qubit
computer is the state \ket{00}. The corresponding density matrix
\begin{equation}
\ket{00}\bra{00}=
\begin{pmatrix}
1 & 0 & 0 & 0 \\ 0 & 0 & 0 & 0 \\ 0 & 0 & 0 & 0 \\ 0 & 0 & 0 & 0
\end{pmatrix}
\label{eq:rho00}
\end{equation}
is not the same as the thermal equilibrium density matrix
\begin{equation}
I_z+S_z=
\begin{pmatrix}
1 & 0 & 0 & 0 \\ 0 & 0 & 0 & 0 \\ 0 & 0 & 0 & 0 \\ 0 & 0 & 0 & -1
\end{pmatrix}.
\end{equation}
The ideal density matrix (Eq.~\ref{eq:rho00}) can, however, be
decomposed as the sum of four product operators:
\begin{equation}\label{eq:pps2}
\ket{00}\bra{00}=\half\left( \half E + I_z + S_z + 2I_zS_z
\right),
\end{equation}
and this sum (ignoring multiples of the unit matrix as usual) can
be assembled using conventional NMR techniques, as described
below.

Superpositions can be treated in much the same way, but they are not directly
related to NMR coherences in any very simple way. For example consider the
state $(\ket{00}+\ket{01})/\sqrt{2}$, in which the first spin is in state
\ket{0}, while the second spin is in a superposition of states.  The
corresponding density matrix can be decomposed directly:
\begin{equation}
\renewcommand{\arraystretch}{1.2}
\begin{pmatrix}
\half & \half & 0 & 0 \\ \half & \half & 0 & 0 \\ 0 & 0 & 0 & 0
\\ 0 & 0 & 0 & 0
\end{pmatrix}
=\half\left( \half E + I_z + S_x + 2I_zS_x \right),
\end{equation}
but there is a more subtle approach.  Note that
$(\ket{00}+\ket{01})/\sqrt{2}$ can be written as a product of
single qubit states
\begin{equation}
\frac{\ket{00}+\ket{01}}{\sqrt{2}}=\frac{\ket{0}(\ket{0}+\ket{1})}{\sqrt{2}},
\end{equation}
and so the corresponding density matrix can also be decomposed as
a direct product of equations \ref{eq:rho0} and \ref{eq:rhox}:
\begin{equation}\begin{split}
\begin{pmatrix} 1 & 0 \\ 0 & 0 \end{pmatrix}
\otimes
\begin{pmatrix} \smhalf & \smhalf \\ \smhalf & \smhalf \end{pmatrix}
&=\left(\half E+I_z\right)\times\left(\half E+S_x\right)\\
&=\half\left(\half E+I_z+S_x+2I_zS_x\right).
\end{split}\end{equation}
Unlike the single qubit case, a simple superposition does not correspond
directly to an NMR coherence, but instead to a complex mixture of coherences
and populations.  It is, however, rarely necessary to worry about this, as
such states can be easily obtained from states like Eq.~\ref{eq:rho00}.

Finally consider superpositions of the form $(\ket{00}+\ket{11})/\sqrt{2}$,
which cannot be broken down into a product of one qubit states (such states
are said to be \emph{entangled}). As they cannot be factored it is necessary
to decompose the corresponding density matrices directly.  In this case
\begin{equation}\label{eq:NMRBell}
\renewcommand{\arraystretch}{1.2}
\begin{pmatrix} \half & 0 & 0 & \half \\ 0 & 0 & 0 & 0 \\ 0 & 0 &
0 & 0 \\ \half & 0 & 0 & \half
\end{pmatrix}
=\half\left(\half E + 2I_zS_z + 2I_xS_x - 2I_yS_y\right),
\end{equation}
which is a mixture of longitudinal two-spin order and $DQ_x$ double quantum
coherence.

\section{NMR logic gates}
\label{sec:NMRgates} After the rather abstract discussions above, we now turn
to the details of methods by which quantum logic gates can be (and have been)
implemented within NMR.

\subsection{One qubit gates}
Many one qubit logic gates can be implemented directly.  For example, a simple
\textsc{not} gate, which interconverts \ket{0} and \ket{1}, can be implemented
as a $180^\circ_x$ rotation \cite{Jones:1998d}. Rotations about axes in the
$xy$-plane can be achieved using RF pulses, while rotations about the $z$-axis
can be accomplished either by using periods of free precession under the
Zeeman Hamiltonian, or by composite $z$-pulses \cite{Freeman:1981}.  This does
not, however, cover the full range of gates which may be desired, as some of
these correspond to rotations about tilted axes.

An obvious (and important) example is the Hadamard gate.  While this
superficially resembles a $90^\circ$ pulse, this resemblance is misleading, as
the Hadamard gate is its own inverse.  Clearly the Hadamard must correspond to
a $180^\circ$ rotation, and a little thought reveals that this rotation occurs
around an axis tilted at $45^\circ$ within the $xz$-plane. This could be
achieved directly by using off-resonance excitation, but this has a number of
practical difficulties.  Alternatively it can be implemented using a composite
pulse sequence, such as $45^\circ_y$--$180^\circ_x$--$45^\circ_{-y}$; as
$180^\circ_x$--$45^\circ_{\pm y}$ can be replaced by $45^\circ_{\mp
y}$--$180^\circ_x$ this three pulse sequence may be simplified to the two
pulse sequence $90^\circ_y$--$180^\circ_x$ or $180^\circ_x$--$90^\circ_{-y}$.

In fact, when implementing quantum algorithms on NMR quantum computers it is
rarely necessary or desirable to use a Hadamard gate, as it can generally be
replaced by the NMR pseudo-Hadamard gate, a $90^\circ_y$ pulse
\cite{Jones:1998d}.  As this gate is not self-inverse it is usually necessary
to replace pairs of Hadamard gates by one pseudo-Hadamard and one inverse
pseudo Hadamard ($90^\circ_{-y}$) gate. This is a simple example of a general
rule in experimental implementations of quantum computation: rather than
directly implementing the gates commonly used in theoretical descriptions, it
is better to use simpler gates which are broadly functionally equivalent to
them.

\subsection{Controlled-NOT gates}
This approach is also applicable to the implementation of controlled two-qubit
gates.  While it is perfectly possible to implement a controlled-\textsc{not}
gate, this is not necessarily the most sensible approach.  The
controlled-\textsc{not} gate can itself be assembled from simpler basic gates,
and it may be more sensible to use these basic gates directly.

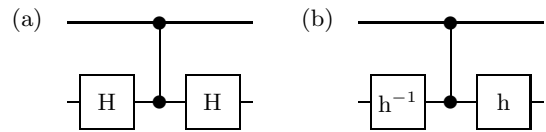
\begin{figure}
\begin{center}
\raisebox{10ex}{(a)}\quad
\begin{picture}(70,45)
\put(0,40){\line(1,0){70}} \put(35,40){\line(0,-1){30}}
\put(35,40){\circle*{5}} \put(35,10){\circle*{5}}
\put(0,10){\line(1,0){5}} \put(5,0){\framebox(20,20){\textsc{H}}}
\put(25,10){\line(1,0){20}}
\put(45,0){\framebox(20,20){\textsc{H}}}
\put(65,10){\line(1,0){5}}
\end{picture}\qquad
\raisebox{10ex}{(b)}\quad
\begin{picture}(70,45)
\put(0,40){\line(1,0){70}} \put(35,40){\line(0,-1){30}}
\put(35,40){\circle*{5}} \put(35,10){\circle*{5}}
\put(0,10){\line(1,0){5}}
\put(5,0){\framebox(20,20){$\text{h}^{-1}$}}
\put(25,10){\line(1,0){20}} \put(45,0){\framebox(20,20){h}}
\put(65,10){\line(1,0){5}}
\end{picture}
\end{center}
\caption{(a) A circuit for implementing a \CNOT\ gate; the two outer gates are
one qubit Hadamard gates, while the central gate is a controlled
$\pi$-phase-shift gate (see text for details). (b) A circuit for implementing
a \CNOT\ gate on an NMR quantum computer; the one qubit Hadamard gates are
replaced by pseudo-Hadamard (h) and inverse pseudo-Hadamard ($\text{h}^{-1}$)
gates.}\label{fig:CNOTn}
\end{figure}

A natural way to implement a \CNOT\ gate is to use a three gate circuit, as
shown in figure \ref{fig:CNOTn}(a).  The two boxes marked \textsc{H} are one
qubit Hadamard gates, and the central gate (two circles connected by a control
line) is a two qubit controlled $\pi$ phase-shift gate. This gate performs the
transformation
\begin{equation}
\ket{1}\ket{1} \overset{\bpi}\longrightarrow -\ket{1}\ket{1}
\end{equation}
while leaving all other states unchanged, and so is described by
the matrix
\begin{equation}
\bpi=
\begin{pmatrix}1&0&0&0\\0&1&0&0\\0&0&1&0\\0&0&0&-1\end{pmatrix}.
\label{eq:cps}
\end{equation}
Unlike the \CNOT\ gate this phase shift gate is symmetric; it is not
meaningful to ask which qubit the phase shift was applied to. Note that the
other \CNOT\ gate, in which the roles of control and target qubit are
reversed, can be constructed by simply moving the two Hadamard gates to the
upper line.

When implementing this circuit on an NMR quantum computer it is preferable to
avoid using Hadamard gates, as these are difficult to implement.  Instead
these two gates are replaced by an inverse pseudo-Hadamard (a $90_{-y}$ pulse)
and a pseudo-Hadamard gate (a $90_y$ pulse) respectively, as shown in figure
\ref{fig:CNOTn}(b).

To see how to implement the controlled phase shift gate it is best to break
down the propagator, equation \ref{eq:cps}, using the product operator basis
set.  As this propagator is diagonal, it must arise from evolution under the
diagonal operators $I_z$, $S_z$, $2I_zS_z$ and (for completeness) $\smhalf E$.
It can be decomposed in two ways,
\begin{equation}
\bpi=\exp\left[-i\times\pi/2\left(\smhalf
E-I_z-S_z+2I_zS_z\right)\right], \label{eq:pi1}
\end{equation}
or
\begin{equation}
\bpi=\exp\left[-i\times\pi/2\left(-\smhalf
E+I_z+S_z-2I_zS_z\right)\right]. \label{eq:pi2}
\end{equation}
The choice between these two decompositions is simply a matter of experimental
convenience.

The use of $\smhalf E$ in these equations might appear to give rise to
difficulties, as this is not normally considered as a product operator which
can arise in NMR Hamiltonians.  In fact it is of no significance at all, as
the effect of the $\smhalf E$ term is simply to impose a global phase shift.
Such global phase shifts have no physical meaning, and cannot be detected;
note that these global phase shifts have no effect on density matrix (or
product operator) descriptions of the spin system. Physically this corresponds
to the fact that there is no absolute zero against which energies may be
measured.

Implementing these propagators (equations \ref{eq:pi1} and \ref{eq:pi2}) is
fairly straightforward. The pure spin--spin coupling term ($2I_zS_z$) can be
generated using conventional spin-echo techniques, while the two Zeeman
Hamiltonians ($I_z$ and $S_z$) can be achieved by appropriately timed periods
of free precession, by the use of composite $z$-pulses, or most simply by just
rotating the RF reference frame. As the three terms all commute, they need not
be applied simultaneously, but can be applied in any order. When the
individual elements making up the propagator are combined, it is frequently
possible to combine or cancel individual pulses, thus simplifying the whole
sequence. This results in a wide variety of possible pulse sequences, and the
choice among them is largely a matter of taste.  For example, one possible
sequence for implementing $\bpi$ is
\begin{equation}
\frac{1}{4J} - 180_x - \frac{1}{4J} - 90_x - 90_{-y} - 90_x
\end{equation}
where all pulses are applied to \emph{both} spins.

While the possible pulse sequences differ in detail they have one
feature in common: an evolution time of $1/2J$ (occasionally
$3/2J$) during which the spins evolve under the spin--spin
coupling, so that the \emph{antiphase} condition is achieved. This
is, of course, the central feature of coherence transfer
sequences, such as INEPT, indicating the close relationship
between controlled two qubit gates and coherence transfer.

\subsection{Other two qubit gates}
While the \CNOT\ gate is important it is not the only two-qubit
gate worth considering: while it is possible to construct any
desired gate using only \CNOT\ gates and one-qubit gates, it is
usually more efficient to use a wider repertoire of basic gates.

One simple and important example is the controlled square-root of \NOT\ gate,
which plays a central role in traditional constructions of the three-qubit
Toffoli gate \cite{DiVincenzo:1998b}; this is one member of a more general
family of $n^{\text{th}}$ roots of \NOT.  Such gates can be built in much the
same way as \CNOT\ gates, except that the controlled $\pi$ phase-shift gate
must be replaced by the more general transformation
\begin{equation}
\ket{1}\ket{1} \overset{\bphi}\longrightarrow
e^{i\phi}\ket{1}\ket{1}
\end{equation}
with $\phi=\pi/n$.  Clearly $\bphi$ can be constructed in much the same way as
$\bpi$, equation \ref{eq:pi2}.

\subsection{Gates in larger spin systems}
The approaches described above can be easily implemented in two spin systems,
allowing quantum computers with two qubits to be easily constructed.  With
larger spin systems, however, the process can become much more complicated
\cite{Jones:2000d}. It is not possible simply to use pulse sequences designed
for two spin systems, as it is necessary to consider the evolution of all the
additional spins in the system.  In particular it may be necessary to refocus
the evolution of these spins under their chemical shift and J-coupling
interactions.  The simplest method is to nest spin echoes within one another,
so that all the undesirable interactions are removed, but this na\"{\i}ve
approach requires an exponentially large number of refocusing pulses (that is,
the complexity of the pulse sequence doubles with every additional spin). This
problem can be overcome by using efficient refocusing sequences
\cite{Jones:1999b, Leung:1999b}, which allow refocusing to be achieved with
quadratic overhead.

It is, of course, rare to find a large spin system where all the
couplings have significant size; in most cases long range
couplings will be small enough to be neglected. This greatly
simplifies the problem, both by reducing the number of couplings
which have to be refocused, and by simplifying the echo sequences
required \cite{Jones:1999b, Linden:1999c}.  It might seem that it
would be difficult to implement some logic gates in such a
partially coupled spin system, as the necessary spin--spin
couplings are missing. In fact this is not a problem, as long as
every pair of spins is connected by some chain of couplings:
quantum \SWAP\ gates \cite{Madi:1998, Linden:1999b} can be used to
move quantum information along this chain.

\subsection{Multi qubit logic gates}
Multi qubit logic gates are gates, such as the Toffoli gate, which perform
controlled operations involving more than two qubits. Such gates can of course
be implemented by constructing appropriate networks of one qubit and two-qubit
gates \cite{DiVincenzo:1998b}, but as the NMR Hamiltonian can contain terms
connecting multiple pairs of spins it should be possible to build some such
gates directly, with a significant saving in pulse sequence complexity.  This
is indeed the case, and several interesting results have been obtained
\cite{Price:1999b, Price:2000a}.

\subsection{Single transition selective pulses}
An alternative approach for building controlled two qubit gates is to use
ultra-soft selective pulses \cite{Freeman:1997}, with excitation profiles so
narrow that they pick out, for example, a single transition from within a
doublet. This corresponds to only exciting the nucleus of interest when the
neighbouring nucleus is in a certain state \cite{Barenco:1995b}; if the
excitation corresponds to a $180^\circ$ pulse then this provides a simple way
of constructing a \CNOT\ gate \cite{Linden:1998a}. The relationship between
this approach and the (more common) multiple pulse sequence approach is
analogous to that between the old fashioned selective population transfer
experiment \cite{Pachler:1973} and its more modern counterpart, INEPT
\cite{Morris:1979}.

One advantage of this approach \cite{Linden:1998a, Dorai:2000} is that it is
relatively easy to extend it to multi qubit gates such as the Toffoli gate.
This can be achieved by using a selective pulse which affects one of the four
transitions of a spin coupled to two neighbours.  A corresponding disadvantage
is that in this case constructing a simple \CNOT\ gate requires either a
selective pulse which excites two of the four transitions in such a system or
the application of two single transition selective pulses in sequence.

The traditional Toffoli gate corresponds to inverting a qubit when two other
qubits are in the state \ket{1}, but there is an entire family of related
gates which effect an inversion for some given pattern of states.  Each such
gate corresponds to exciting a different transition in the multiplet, and so
the entire family of gates can be achieved directly.  In practice, however,
the central regions of a multiplet can become quite crowded, with many lines
nearly overlapping, and it will be difficult to select a single transition.
By contrast the two transitions at the extreme ends of the multiplet will
always be relatively well separated from their nearest neighbours, and it is
best to concentrate on these two frequencies.  Other gates can then be
constructed by surrounding these basic gates with \NOT\ gates applied to the
neighbouring spins, thus permuting the identities of the lines in the
multiplet.

\subsection{Geometric phase-shift gates}
A third approach for implementing NMR quantum computation, based on the use of
geometric phase-shift gates, has recently been described \cite{Jones:2000a,
Ekert:2000a}. Like the conventional approach it relies on controlled
phase-shift gates, but the phase shifts are generated using geometric phases
\cite{Shapere:1989}, such as Berry's phase \cite{Berry:1984}, rather than the
more conventional dynamic phases.  Berry phases have been demonstrated in a
wide variety of systems \cite{Shapere:1989}, including NMR \cite{Suter:1987,
Goldman:1996} and the closely related technique of NQR \cite{Tycko:1987,
Appelt:1994, Jones:1997}, and can be used to implement controlled phase shift
gates in NMR systems \cite{Jones:2000a, Ekert:2000a}. This approach has few
advantages for NMR quantum computation, but may prove useful in other systems
\cite{Ekert:2000a}.

\section{Initialisation and NMR}\label{sec:NMRinit}
As it is impractical to cool down NMR spin systems to their ground state
\cite{Jones:2000d, Warren:1997, Gershenfeld:1997b}, initialisation of an NMR
quantum computer in practice means assembling an appropriate pseudo-pure
state.  This approach is useful only if some practical procedure for
assembling such states can be devised.

For the simplest possible system (a single nucleus) the process is trivial, as
the thermal equilibrium density matrix has the desired form, but with larger
systems the situation is more complicated. The essential feature of a
pseudo-pure state is that it has a diagonal density matrix in which the
populations of all the spin states (the elements along the diagonal) are the
same, with the exception of one state (normally $\ket{\mathbf{0}}=
\ket{000\dots 0}$) which has a larger population.  By contrast, at thermal
equilibrium the spin state populations are distributed in accordance with the
Boltzmann equation, and so exhibit a more complex variation.  For a
homonuclear two spin system the equilibrium density matrix (neglecting
multiples of the identity matrix and an initial scaling factor) is $I_z+S_z$,
while the desired pseudo-pure state is proportional to $I_z + S_z + 2I_zS_z$
(equation \ref{eq:pps2}).

The original approach for assembling pseudo-pure states, developed by Cory
\emph{et al.} \cite{Cory:1996, Cory:1997}, uses conventional NMR techniques.
Assembling such a mixture using pulse sequences and field gradients is a
fairly straightforward, if somewhat unusual, NMR problem.  This process is
commonly called \emph{spatial averaging}, presumably a reference to the use of
field gradients.

A second early approach, suggested by Gershenfeld and Chuang
\cite{Gershenfeld:1997}, is to use a subset of the energy levels in a more
complex spin system. For example, in a homonuclear three spin system it is
possible to find a set of four energy levels which exhibit the pattern of
populations corresponding to the pseudo-pure state of a two spin system.  This
approach, often called \emph{logical labeling}, is elegant in principle but
complex to apply in practice, and has only rarely been experimentally
demonstrated.

More recently a variety of different approaches have been used, although these
all combine elements of the two basic approaches above.  The most popular
technique, usually called \emph{temporal averaging} \cite{Knill:1998}, works
by performing many different experiments, each with a different initial state.
For example, in a two qubit system, one might perform experiments starting
from $I_z$, $S_z$, and $2I_zS_z$.  If the spectra from these three experiments
are then added together, the result is equivalent to a single experiment
starting from a mixture of these states.  Clearly temporal averaging and
spatial averaging are related in much the same way as coherence selection
methods based on phase cycling and gradients.  Finally, a new approach
combines these methods in a cunning way, using the analogy between multiple
quantum coherence and so-called ``cat'' states to generate pseudo-pure states
in a fairly efficient manner \cite{Knill:2000}.

\subsection{Spatial averaging}
The direct ``spatial averaging'' technique may be exemplified by the original
sequence of Cory \emph{et al.} \cite{Cory:1996, Cory:1997} for constructing a
pseudo-pure state in a two spin system:
\begin{equation}
\begin{split}
&I_z+S_z\\
\xrightarrow{\makebox[2.5em]{$\scriptstyle60^{\circ}S_x$}}
&I_z+{\smhalf}S_z-\raisebox{0.4ex}{$\scriptstyle{\frac{\sqrt3}{2}}$}S_y\\
\xrightarrow{\makebox[2.5em]{$\scriptstyle\text{crush}$}}
&I_z+{\smhalf}S_z\\
\xrightarrow{\makebox[2.5em]{$\scriptstyle45^{\circ}I_{x}$}}
&\raisebox{0.4ex}{$\scriptstyle{\frac{1}{\sqrt2}}$}I_z
-\raisebox{0.4ex}{$\scriptstyle{\frac{1}{\sqrt2}}$}I_y+{\smhalf}S_z\\
\xrightarrow{\makebox[2.5em]{$\scriptstyle\text{couple}$}}
&\raisebox{0.4ex}{$\scriptstyle{\frac{1}{\sqrt2}}$}I_z
+\raisebox{0.4ex}{$\scriptstyle{\frac{1}{\sqrt2}}$}2I_xS_z+{\smhalf}S_z\\
\xrightarrow{\makebox[2.5em]{$\scriptstyle45^{\circ}I_{-y}$}}
&{\smhalf}I_z-{\smhalf}I_x+{\smhalf}2I_xS_z+{\smhalf}S_z+{\smhalf}2I_zS_z\\
\xrightarrow{\makebox[2.5em]{$\scriptstyle\text{crush}$}}
&{\smhalf}I_z+{\smhalf}S_z+{\smhalf}2I_zS_z\\
\end{split}
\end{equation}
where the sequence is described in product operator notation, ``crush''
indicates the application of a crush field gradient pulse, and ``couple''
indicates evolution under the scalar coupling for a time $1/2J$. Note that the
two crush pulses must be applied along different axes, or with different
strengths, to prevent undesired terms from being refocused.

Several alternative sequences for creating two spin pseudo-pure states have
been developed; for example \cite{Pravia:1999}
\begin{equation}
\begin{split}
I_z+S_z &\xrightarrow{45^{\circ}(I_x+S_x)}
\xrightarrow{\text{couple}}
\xrightarrow{30^{\circ}(I_{-y}+S_{-y})}\\
&\xrightarrow{\text{crush}}
\sqrt{\frac{3}{8}}\left(I_z+S_z+2I_zS_z\right)
\end{split}
\end{equation}
where zero quantum terms (which in a homonuclear spin system will survive the
crush pulse) have been neglected.  This scheme works well in heteronuclear
spin systems, but in homonuclear systems it is necessary to use a more complex
approach to deal with the zero quantum terms.

These sequences can be generalised to larger spin systems, but this process is
quite complex.  For this reason most work on larger spin systems has used
temporal averaging techniques. Recently, however, Knill \emph{et al.}
\cite{Knill:2000} have developed a general scheme based on cat states, which
allows pulse sequences for any spin system to be developed.  This approach is
described below.

\subsection{Logical labeling}
Logical labeling \cite{Gershenfeld:1997} is most easily understood by
examining the thermal equilibrium density matrix for a homonuclear three spin
system:
\begin{equation}
I_z + S_z + R_z = \half\left\{3, 1, 1, -1, 1, -1, -1, -3 \right\},
\end{equation}
where the braces indicate a diagonal matrix defined by listing its diagonal
elements.  While this matrix does not have the right form for a three spin
pseudo-pure state it is possible to select out four levels (corresponding to
the states \ket{000}, \ket{011}, \ket{101} and \ket{110}) which have the same
population pattern as
\begin{equation}
I_z+S_z+2I_zS_z=\half\left\{3,-1,-1,-1\right\},
\end{equation}
and so this subset of levels can be used as a two spin pseudo-pure state.

It would be possible to use these states directly, but this would greatly
complicate subsequent logic operations as there is no simple correspondence
between these four states of the three spin system and the four basic states
of a two spin system.  Instead it is better to permute the populations of the
various states, performing $\ket{001}\leftrightarrow\ket{101}$ and
$\ket{010}\leftrightarrow\ket{110}$; these permutations can be achieved using
\CNOT\ gates.  At the end of this process the populations are given by
\begin{equation}
\half\left\{3,-1,-1,-1,1,1,1,-3\right\},
\end{equation}
so that the states \ket{000}, \ket{001}, \ket{010} and \ket{011}
are in a pseudo-pure state.  Note that these four states all have
the first spin in state \ket{0}, and so the first spin acts as an
ancilla spin, labeling the ``correct'' subspace.

Similar, but more complex, procedures can be used with larger spin systems
\cite{Gershenfeld:1997}. The overhead required is fairly small; that is the
number of pseudo-pure spins which can be encoded in a spin system is only
slightly smaller than the size of the system.  However, while these results
are elegant the complexity of implementing logical labeling means that
experimental demonstrations have so far been confined to three spin systems
\cite{Dorai:2000, Vandersypen:1999}.

\subsection{Temporal averaging}
As discussed previously, temporal averaging \cite{Knill:1998} bears much the
same relationship to spatial averaging as phase cycling does to the use of
gradients to select coherence transfer pathways.  The name can, however, be
used to cover a variety of different approaches.

As described above (equation \ref{eq:pps2}), a pseudo-pure state of a two spin
system can be assembled as a mixture of three terms: $I_z$, $S_z$ and
$2I_zS_z$.  The simplest approach to temporal averaging is just to perform a
computation starting from each of these states, and add the results together
at the end.  This is easily generalised to larger spin systems: for a system
of $n$ spins it is necessary to perform $2^{n}-1$ separate experiments. In
some simple experiments it is possible to show that only some of these
starting states will give an observable signal \cite{Marx:2000}, and so it is
unnecessary to perform experiments starting in other states. This permits
substantial experimental simplifications, but it is not a general technique.

A better approach is to use the original scheme of Knill \emph{et al.}
\cite{Knill:1998}. The thermal equilibrium density matrix for a two spin
system
\begin{equation}
I_z+S_z=\left\{1, 0, 0, -1\right\}
\end{equation}
(where the braces have the same meaning as before) can be easily converted
into two other states,
\begin{equation}
I_z+2I_zS_z=\left\{1, 0, -1, 0\right\}
\end{equation}
and
\begin{equation}
S_z+2I_zS_z=\left\{1, -1, 0, 0\right\}.
\end{equation}
These three states are related by simple permutations of the populations of
the levels, which can be achieved using \CNOT\ gates.  Adding together the
three starting states gives
\begin{equation}
2\left(I_z+S_z+2I_zS_z\right)=\left\{3, -1, -1, -1\right\},
\end{equation}
which is a pseudo-pure state.  Adding together the spectra from computations
started in these three states therefore gives the spectrum which would be
produced from a pseudo-pure state.

Once again this process is easily generalised to larger spin systems.  The
most obvious approach is to average over the $2^n-1$ cyclic permutations of
the populations in an $n$ spin system. This \emph{exhaustive averaging} scheme
is just as inefficient as the na\"{\i}ve approach outlined above, but Knill
\emph{et al.} \cite{Knill:1998} have shown that similar results can be
achieved by averaging over much smaller numbers of states.

\subsection{The use of ``cat'' states}
The schemes described above are perfectly practical for small spin systems but
are harder to use with larger systems.  Recently Knill \emph{et al.}
\cite{Knill:2000} have described a simple approach which works for spin
systems of any size and which can be used with either the gradient (spatial
averaging) or phase cycling (temporal averaging) approaches.  Their method is
based on the properties of ``cat'' states, named by analogy with
Schr\"odinger's Cat. An $n$ qubit cat state is a superposition state of the
form
\begin{equation}
\phi_n^\pm=\left(\ket{00\dots 0}\pm\ket{11\dots
1}\right)/\sqrt{2},
\end{equation}
so that either all the $n$ qubits are in state \ket{0}, or all the qubits are
in state \ket{1}.  (In fact the relative phase of the two states contributing
to the superposition can take any value between $0$ and $2\pi$, but it is
convenient to restrict ourselves to the two values $0$ and $\pi$, giving rise
to the factor of $\pm 1$.)  States of this form are said to be
\emph{entangled}, and play a central role in quantum information processing
and experimental tests of quantum mechanics.  The role of entanglement in NMR
quantum computers will be explored in more detail below, but for the moment it
is sufficient to note that cat states are closely related to (but not simply
equivalent to) multiple quantum coherence \cite{Jones:1998d}.

As discussed above (equation \ref{eq:NMRBell}) the two qubit cat state
$\phi_2^+=(\ket{00}+\ket{11})/\sqrt{2}$ (commonly called a Bell state
\cite{Bouwmeester:2000b}) is a mixture of $DQ_x$ double quantum coherence and
longitudinal two-spin order. Similarly the three qubit cat state $\phi_3^+$
(usually called a GHZ state \cite{Bouwmeester:2000b}) is a mixture of $3Q_x$
triple quantum coherence and the three possible states of longitudinal
two-spin order, and a general $n$ qubit cat state will correspond to a mixture
of $n$ quantum coherence and ordered population states. Thus an $n$ quantum
filtration sequence is almost (though not quite) equivalent to selecting $n$
qubit cat states.

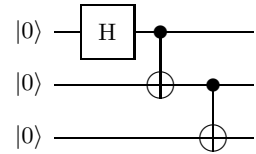
\begin{figure}
\begin{center}
\begin{picture}(120,55)
\put(20,50){\line(1,0){10}}
\put(30,40){\framebox(20,20){\text{H}}}
\put(50,50){\line(1,0){45}}
\put(60,50){\line(0,-1){25}}\put(60,50){\circle*{5}}
\put(20,30){\line(1,0){75}}\put(60,30){\circle{10}}
\put(80,30){\line(0,-1){25}}\put(80,30){\circle*{5}}
\put(20,10){\line(1,0){75}}\put(80,10){\circle{10}}
\put(5,48){\ket{0}}\put(5,28){\ket{0}}\put(5,8){\ket{0}}
\end{picture}
\end{center}
\caption{A network for creating the three qubit cat state
$\phi_3^+$.} \label{fig:cat}
\end{figure}

Cat states are easily prepared from pure states, using \CNOT\ gates.  One
possible network for a three qubit system is shown in figure \ref{fig:cat};
networks for larger systems can be derived by analogy.  Similarly by reversing
this network cat states can be converted back into pure states.  Thus, if it
is possible to prepare an $n$ qubit cat state, it should be possible to obtain
a corresponding pure state.

This suggests a simple scheme for preparing pseudo-pure states. If the network
shown in figure \ref{fig:cat} is applied to a three spin system in its thermal
equilibrium state, the resulting mixture will include a component of triple
quantum coherence, and thus of the desired cat state. This component can be
selected, either by phase cycling or by using gradient methods.  Finally the
network can be reversed to convert the cat state back into a pseudo-pure
state.

Unfortunately this does not quite have the desired effect, as triple quantum
coherence is not quite equivalent to the desired cat state; in fact $3Q_x$
corresponds to $\ket{\phi_3^+}\bra{\phi_3^+}-\ket{\phi_3^-}\bra{\phi_3^-}$,
and so \emph{both} cat states will be retained by the triple quantum filter.
The effect of reversing the network is then to convert this to the pseudo-pure
state corresponding to
\begin{equation}\label{eq:ppcat}
\ket{000}\bra{000}-\ket{100}\bra{100}=I_z\otimes\ket{00}\bra{00}.
\end{equation}
This \emph{is} a pseudo-pure state of the last two spins, and in general
multiple quantum selection of cat states provides a convenient way of
generating an $n-1$ qubit pseudo-pure state in an $n$ spin system.
Furthermore, for some purposes states of the form given by equation
\ref{eq:ppcat} can be used as if they were $n$ qubit pseudo-pure states
\cite{Knill:2000}.

\section{Readout}
As described above there are two main methods for determining the final state
of an NMR quantum computer: by examining the spectrum of the spins
corresponding to the qubits of interest, and by examining the spectra of other
neighbouring spins.  These methods are simplest to describe when the quantum
computer ends its computation with all the answer qubits in the eigenstates
\ket{0} and \ket{1}, rather than in superposition states or entangled states,
as in this case a small number of measurements will provide all the
information required \cite{Jones:1998a}.  A more thorough approach is to
completely characterise the final state of the spin system by so called
quantum state tomography \cite{Chuang:1998a}; while the results can be
interesting in small spin systems the effort required to perform tomography
increases rapidly with the size of the spin system, and this approach is
probably impractical for systems of more than three spins.

\subsection{Simple readout}
The simplest situation to consider is a one qubit NMR quantum computer which
ends a calculation in the pseudo-pure state corresponding to \ket{0} or
\ket{1}.  As discussed above (equations \ref{eq:rho0} and \ref{eq:rho1}),
these correspond to the NMR states $I_z$ and $-I_z$ respectively, and
excitation with a $90^\circ\, I_y$ pulse will convert these to $\pm I_x$. Thus
the two states will give rise to absorption and emission lines in the NMR
spectrum; this is hardly surprising as they correspond to excess population in
the low energy and high energy spin states. It is, of course, necessary to
obtain a reference signal against which the phase of the signal of interest
can be determined, but this is easily achieved, either by using the NMR signal
from a reference compound, or by acquiring a signal from the computer in a
known state, \ket{0} or \ket{1}.

The situation is similar, but more complex, with larger spin systems.  The NMR
state corresponding to \ket{00} is not just $I_z+S_z$, as might na\"{\i}vely
be expected; instead it is $I_z+S_z+2I_zS_z$ (see equation \ref{eq:pps2}).  A
general two qubit pseudo-pure eigenstate can be expressed similarly as
\begin{equation}\label{eq:rhoab}
\ket{ab}\bra{ab}=\half\left((-1)^a I_z + (-1)^b S_z + (-1)^{a\oplus b}
2I_zS_z\right).
\end{equation}
This can be analysed in two ways: by exciting and observing both spins, or by
exciting and observing just one spin, say $I$.  The first approach is perhaps
the most natural approach in a homonuclear spin system, while the second
method is more appropriate in a heteronuclear spin system.

If both spins are excited, then the two population terms ($I_z$ and $S_z$) are
converted to single quantum coherences, while the longitudinal two spin order
is converted to unobservable double and zero quantum coherence.  Thus the
observable signal from a state of the form equation \ref{eq:rhoab} is
proportional to
\begin{equation}
(-1)^a I_x + (-1)^b S_x.
\end{equation}
Clearly the desired information can be obtained from the phases (absorption or
emission) of the NMR signals from the two spins.

The situation is slightly more complicated if only one spin is observed:
application of a $90^\circ\, I_y$ pulse to the state equation \ref{eq:rhoab}
gives
\begin{equation}
\left((-1)^a I_x + (-1)^b S_z + (-1)^{a\oplus b} 2I_xS_z\right)/2,
\end{equation}
and the observable signal is proportional to
\begin{equation}
(-1)^a \left( I_x + (-1)^b 2I_xS_z\right).
\end{equation}
Thus only one of the two lines in the $I$ spin doublet will be observed; which
of the two lines this is depends on $b$, the state of spin $S$, while the
phase of the signal depends on $a$, the state of spin $I$, as before.

\subsection{Tomography}
Many NMR quantum computation experiments have used a readout scheme called
quantum state tomography, and while this scheme is impractical for use with
large spin systems it merits some explanation.  The easiest approach to
readout is simply to determine the states of one or more critical qubits which
contain the desired answer, while an alternative, far more thorough, approach
is to characterise the complete density matrix describing the final state of
the system \cite{Chuang:1998a}. This state tomography approach requires a
large number of different measurements to fully characterise all the elements
of density matrix, and for large spin systems the complexity of this approach
becomes prohibitive.  For small systems, however, it provides detailed
information not just on the result of the calculation, but also on any error
terms.

The density matrix describing a two spin system can itself be described using
fifteen real numbers, corresponding to the amounts of the fifteen two-spin
product operators in the state (neglecting the identity matrix as usual).  In
a heteronuclear spin system it is possible to determine the values of four of
these coefficients (the amounts of $I_x$, $I_y$, $2I_xS_z$, and $2I_yS_z$)
just by observing the $I$ spin free induction decay, while four more can be
determined by observing spin $S$.  The seven remaining coefficients can then
be determined in a minimum of two more experiments by exciting either $I$ or
$S$ before observation.  In general the spectrum of a single spin can provide
at most $2^n$ real numbers, while $4^n-1$ numbers are required to characterise
the spin system; thus a minimum of $2^n$ separate experiments will be
required.  In practice the schemes actually used are substantially less
efficient, greatly increasing the effort required for full tomography.  For
example, one tomographic analysis of a heteronuclear two qubit system involved
nine separate experiments \cite{Chuang:1998b}.

\section{Practicalities}

\subsection{Selective pulses}
Implementing these pulse sequences in a fully heteronuclear spin system is
straightforward, but in a homonuclear spin system complications arise from the
need to perform selective excitation. The simplest approach is the use of
conventional selective pulses \cite{Freeman:1997}. These pulses can be simple
Gaussian pulses incorporating a phase ramp to allow off-resonance excitation,
but it is probably better to use more subtle pulse shapes, such as members of
the BURP family of pulses \cite{Linden:1999b}. The soft pulses should excite
all the lines in the target multiplet in an identical fashion, while leaving
other lines completely untouched. In practice this is difficult to achieve in
\nuc{1}{H} systems, leading to the substantial errors clearly visible in many
experiments.

As pulse sequences implementing quantum logic gates can contain a large number
of selective pulses separated by delays, it is necessary to address each spin
in its own rotating frame. In homonuclear two-spin systems, however, such as
those used to implement two qubit NMR computers, it is possible to use a
simpler approach.  Suppose the centres of the two multiplets are separated by
$\rm\nu\,Hz$; in this case the two frames will rotate with a relative
frequency $\nu$.  If the rotating frames were aligned at the beginning of the
pulse sequence, they will come back into alignment at time intervals $1/\nu$.
As long as excitation and observation is performed stroboscopically it is
possible to treat both nuclei as inhabiting the same rotating frame.
Similarly, by choosing times such that the two rotating frames are $90$ or
$180^\circ$ out of phase, it is possible to use variations on the simple
``jump and return'' pulse sequence \cite{Plateau:1982} to perform selective
excitation.  This approach \cite{Jones:1999a} can prove simpler than using
selective pulses directly, but it cannot easily be used in systems with more
than two spins of a given nuclear species.

\subsection{Composite Pulses}
Composite pulses \cite{Freeman:1997, Levitt:1986} play an important role in
many NMR experiments, enabling the effects of experimental imperfections, such
as pulse length errors and off-resonance effects, to be reduced.  Such pulses
could also prove useful in NMR quantum computers, acting to reduce systematic
errors in quantum logic gates \cite{Cummins:2000a}. Unfortunately most
conventional composite pulse sequences are not appropriate for quantum
computers as they only perform well for certain initial states, while pulse
sequences designed for quantum information processing must act as
\emph{general rotors}, that is they must perform well for \emph{any} initial
state.

Composite pulses of this kind (sometimes called Class A composite pulses
\cite{Levitt:1986}) are rarely if ever needed for more conventional NMR
experiments, and so have been relatively little studied.  One important
example is a composite $90^\circ$ pulse developed by Tycko \cite{Levitt:1986,
Tycko:1983}, which has recently been generalised to arbitrary rotation angles
\cite{Cummins:2000a}. These composite pulses give excellent compensation of
off-resonance effects at small offset frequencies, such as those found for
\nuc{1}{H} nuclei, but are of no use for the much larger off-resonance
frequencies typically found for \nuc{13}{C}.

Fortunately when composite pulses are used for NMR quantum computation one
great simplification can be made: it is only necessary that the pulse sequence
perform well over a small number of discrete frequency ranges, corresponding
to the resonance frequencies of the nuclei used to implement qubits; it is
\emph{not} necessary to design pulses which work well over a broad frequency
range. In particular many NMR quantum computers  use at most two spins of each
nuclear species (see, for example, \cite{Marx:2000}), and it is convenient to
place the RF frequency in the centre of the spectrum, so that the two spins
have equal and opposite resonance offsets \cite{Jones:1999a}. Thus it is
sufficient to tailor the composite pulse sequence to work well at these two
frequencies, while the performance at all other frequencies can be completely
ignored \cite{Cummins:2000b}.

\subsection{Abstract reference frames}
One technique which has proved extremely useful in the implementation of NMR
quantum computers with more than two qubits is the use of \emph{abstract
reference frames} \cite{Knill:2000}.  As it is necessary to address each spin
in its own rotating frame of reference, it is possible to simply rotate this
frame to absorb the effects of $z$ rotations, whether these arise from
attempts to implement quantum gates (see equations \ref{eq:pi1} and
\ref{eq:pi2}), or the failure to fully refocus chemical shifts.

For example, $90^\circ_{\pm z}$ rotations occur in the implementations of many
interesting gates.  If need be these can be achieved either by periods of free
precession, or by composite $z$-pulses.  A simpler approach, however, is to
achieve the same effect by rotating the RF reference frame, so that subsequent
pulses are applied with appropriate phase shifts.  Thus, for example, the
pulse sequence $90_z \, 90_x$ can be replaced by $90_{-y} \, 90_z$: the phase
of the RF pulse has been shifted, and the $z$-pulse has been delayed.  Ideally
it is possible to use this method to delay the $z$ rotation to the very end of
the pulse sequence, where it can be replaced by a rotation of the RF detection
axis, or in many cases ignored all together.

\section{Simple algorithms}\label{sec:algs}
Now that we have seen all the elements necessary to implement quantum logic
operations within NMR it is useful to see how they can be assembled to build
small NMR quantum computers.  Only two algorithms will be discussed in detail,
both of which can be implemented using two qubit computers, that is two spin
systems. Brief reference will, however, be made to more complex systems.

Computers as small as these bear little immediate resemblance to the computers
in widespread use today: with only two qubits there is simply no available
memory in which to store extraneous data or programs!  Instead the program is
built into the design of the NMR pulse sequence used to implement the
computation, and the two qubits are used to store the input data and the
result of the computation, as well as forming the ``CPU'' of the system.

\subsection{Functions and phases}
Before discussing the algorithms themselves, it is useful to describe a trick
widely used in quantum computation for converting the results of a function
evaluation into a phase shift.  This phase trick plays a central role in
conventional implementations of many algorithms, but with NMR quantum
computers it is often more appropriate to redesign the computer to implement
the desired phase shifts directly.

\begin{figure}
\begin{center}
\begin{picture}(80,55)
\put(20,40){\line(1,0){10}} \put(40,30){\line(0,-1){25}}
\put(30,30){{\framebox(20,20){$f$}}} \put(50,40){\line(1,0){10}}
\put(20,10){\line(1,0){40}} \put(40,10){\circle{10}} \put(5,38){$\ket{x}$}
\put(5,8){$\ket{0}$} \put(65,38){$\ket{x}$} \put(65,8){$\ket{f(x)}$}
\end{picture}
\end{center}
\caption{A quantum circuit for the classical analysis of a binary function
$f$; this circuit is drawn for a quantum computer but is equivalent to that
for a classical reversible computer (see figure \ref{fig:fCNOT}).}
\label{fig:f01}
\end{figure}
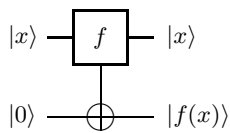
These simple demonstration algorithms are based on the analysis of one-bit
binary functions, that is functions which take in one or more bits as input
and return a single bit (that is, 0 or 1) as output.  These functions can be
evaluated on reversible computers using $f$-\CNOT\ gates, as shown in figure
\ref{fig:fCNOT}, with the result returned as the value of an additional output
bit, which begins the computation initialised to 0.  An equivalent approach
can be used with quantum computers, figure \ref{fig:f01}, but it is also
possible to perform function evaluation with this ``output'' qubit set not to
\ket{0} but to the superposition $(\ket{0}-\ket{1})/\sqrt{2}$. Since
\begin{equation}
\frac{\ket{0\oplus b}-\ket{1\oplus b}}{\sqrt{2}}= (-1)^b
\frac{\ket{0}-\ket{1}}{\sqrt{2}}
\end{equation}
an $f$-\CNOT\ will perform the transformation
\begin{equation}\label{eq:fpt}
\frac{\ket{x}\left(\ket{0}-\ket{1}\right)}{\sqrt{2}}
\overset{f}{\longrightarrow}
(-1)^{f(x)}\frac{\ket{x}\left(\ket{0}-\ket{1}\right)}{\sqrt{2}}
\end{equation}
and so the result of the function is returned as a phase.  Note that the
starting state of the ancilla qubit can be easily prepared from the state
\ket{1} by the application of a Hadamard gate (equation \ref{eq:H1}).

This phase trick might seem pointless, indeed counterproductive, as it seems
to return the result of the function as a global phase, and such global phases
have no physical meaning.  As we shall see, however, the phase trick can be
combined with quantum parallelism in a cunning and useful way.

\subsection{Deutsch's algorithm}
Deutsch's algorithm \cite{Deutsch:1986, Deutsch:1992, Cleve:1998} is concerned
with the analysis of binary functions from one bit to one bit, that is
functions which take in one bit as input and return another bit as output.
Clearly there are four such functions, as shown in table \ref{tab:f1bit}.
\begin{table}
\begin{center}
\begin{tabular}{|l|l|l|l|l|}
\hline $x$&$f_{00}(x)$&$f_{01}(x)$&$f_{10}(x)$&$f_{11}(x)$
\\\hline 0& 0 & 0 & 1 & 1\\ 1& 0 & 1 & 0 & 1\\\hline
\end{tabular}
\end{center}
\caption{The four possible binary functions mapping one bit to
another; each function is conveniently labeled by the bit pattern
in its truth table.} \label{tab:f1bit}
\end{table}
These four functions can be divided into two groups: the two \emph{constant}
functions, for which $f(x)$ is independent of $x$ ($f_{00}$ and $f_{11}$), and
the two \emph{balanced} functions, for which $f(x)$ is zero for one value of
$x$ and one for the other ($f_{01}$ and $f_{10}$).  Equivalently, the
functions can be classified according to the parity of the function,
$f(0)\oplus f(1)$.

Given some unknown function $f$ (chosen from among these four functions), it
is possible to determine which function it is by applying $f$ to two inputs,
$0$ and $1$, using the circuit shown in figure \ref{fig:f01}. This procedure
also provides enough information to determine the parity of $f$, and thus
whether the function is constant or balanced. However knowing the parity of
$f$ corresponds to only one bit of information, and so it might be possible to
answer this question using only one evaluation of the function $f$.  This
cannot be achieved with a classical computer, but with a quantum computer the
problem can be solved using Deutsch's algorithm.

The basic idea behind Deutsch's algorithm is to combine the phase trick with
quantum parallelism.  Suppose that the $f$-\CNOT\ gate is applied with the
input qubit in the state $(\ket{0}+\ket{1})/\sqrt{2}$ and the ancilla qubit in
the state $(\ket{0}-\ket{1})/\sqrt{2}$; then from equation \ref{eq:fpt} the
result of the computation will be
\begin{equation}
\begin{split}
&\left(\frac{(-1)^{f(0)}\ket{0}+(-1)^{f(1)}\ket{1}}{\sqrt{2}}\right)
\left(\frac{\ket{0}-\ket{1}}{\sqrt{2}}\right)\\ =&
(-1)^{f(0)}\left(\frac{\left(\ket{0}+(-1)^{f(0)\oplus
f(1)}\ket{1}\right) \left(\ket{0}-\ket{1}\right)}{2}\right).
\end{split}
\label{eq:fd}
\end{equation}
The ancilla qubit remains in $(\ket{0}-\ket{1})/\sqrt{2}$, while the ``input''
qubit now contains the state $(\ket{0}\pm\ket{1})/\sqrt{2}$, where the choice
of plus or minus sign depends on $f(0)\oplus f(1)$.  As relative phases in
superpositions \emph{can} be detected (for example, by applying a Hadamard
gate as shown in equations \ref{eq:H0} and \ref{eq:H1}), this allows the
parity of $f$ to be determined with only one function evaluation.

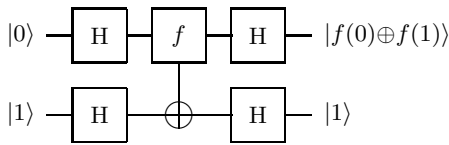
\begin{figure}
\begin{center}
\begin{picture}(170,55)
\put(20,40){\line(1,0){10}}\put(30,30){{\framebox(20,20){\text{H}}}}
\put(50,40){\line(1,0){10}} \put(60,30){{\framebox(20,20){$f$}}}
\put(80,40){\line(1,0){10}}\put(90,30){{\framebox(20,20){\text{H}}}}
\put(110,40){\line(1,0){10}} \put(70,30){\line(0,-1){25}}
\put(20,10){\line(1,0){10}}\put(30,0){{\framebox(20,20){\text{H}}}}
\put(50,10){\line(1,0){40}}\put(70,10){\circle{10}}
\put(90,0){{\framebox(20,20){\text{H}}}}\put(110,10){\line(1,0){10}}
\put(5,38){$\ket{0}$}\put(5,8){$\ket{1}$}
\put(125,38){$\ket{f(0){\oplus}f(1)}$}\put(125,8){$\ket{1}$}
\end{picture}
\end{center}
\caption{A quantum circuit implementing Deutsch's algorithm to determine the
parity of a binary function $f$.} \label{fig:fd}
\end{figure}

A quantum circuit for Deutsch's algorithm is shown in figure \ref{fig:fd}. The
Hadamard gates act to interconvert eigenstates and superpositions, allowing
both the phase trick and quantum parallelism to be implemented.  Note that in
this algorithm there is no input, and the result ends up in the first qubit,
not the second qubit as occurs for traditional function evaluation. The second
qubit is used simply as an ancilla to implement the phase trick. As discussed
previously , it is not possible to program such a simple computer; instead the
choice of function $f$ is ``hard wired'' into the computer by the design of
the $f$-\CNOT\ gate.

\subsection{NMR implementations}
Deutsch's algorithm was not only the first quantum algorithm to be described:
it was also the first algorithm to be implemented on an NMR quantum computer,
first using a homonuclear system (cytosine) and then a heteronuclear system
(\nuc{13}{C}-labeled chloroform). These two implementations will be described
below, while more modern implementations, including some extensions and
simplifications, are described in the next section.

The cytosine system \cite{Jones:1998a} used the two  \nuc{1}{H} nuclei
remaining on a cytosine molecule when dissolved in $\rm D_2O$ (figure
\ref{fig:cytosine}). The two \nuc{1}{H} multiplets are separated by
$\rm763\,Hz$ (at a \nuc{1}{H} frequency of $\rm500\,MHz$), with a J-coupling
of $\rm7.2\,Hz$.  Selective excitation was achieved using Gaussian shaped soft
pulses incorporating a phase ramp, and the lengths of the soft pulses were
chosen as multiples of the inverse of the frequency separation of the two
resonances, so that the unexcited spin experienced no net rotation during a
selective pulse.

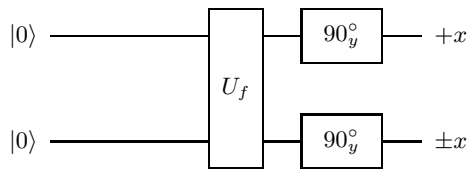
\begin{figure}
\begin{center}
\begin{picture}(200,70)
\put(15,50){\makebox(0,0)[r]{\ket{0}}} \put(20,50){\line(1,0){60}}
\put(15,10){\makebox(0,0)[r]{\ket{0}}} \put(20,10){\line(1,0){60}}
\put(80,0){\framebox(20,60){$U_f$}} \put(100,50){\line(1,0){15}}
\put(115,40){\framebox(30,20){$90^\circ_{y}$}}
\put(145,50){\line(1,0){15}} \put(165,50){\makebox(0,0)[l]{$+x$}}
\put(100,10){\line(1,0){15}}
\put(115,0){\framebox(30,20){$90^\circ_{y}$}}
\put(145,10){\line(1,0){15}} \put(165,10){\makebox(0,0)[l]{$\pm
x$}}
\end{picture}
\end{center}
\caption{Modified quantum circuit for the classical analysis of $f(0)$ on an
NMR quantum computer; $U_f$ is a propagator corresponding to the $f$-\CNOT\
gate in the conventional circuit (figure \ref{fig:f01}).  Function evaluation
is followed by $90^\circ_y$ pulses to excite the NMR spectrum.  Clearly $f(1)$
can be obtained in a very similar fashion.} \label{fig:f01nmr}
\end{figure}

\begin{figure}
\begin{center}
\begin{picture}(200,70)
\put(15,50){\makebox(0,0)[r]{\ket{0}}} \put(20,50){\line(1,0){15}}
\put(35,40){\framebox(30,20){$90^\circ_y$}}
\put(65,50){\line(1,0){15}} \put(15,10){\makebox(0,0)[r]{\ket{1}}}
\put(20,10){\line(1,0){15}}
\put(35,0){\framebox(30,20){$90^\circ_y$}}
\put(65,10){\line(1,0){15}} \put(80,0){\framebox(20,60){$U_f$}}
\put(100,50){\line(1,0){60}} \put(165,50){\makebox(0,0)[l]{$\pm
x$}}
\put(100,10){\line(1,0){60}}\put(165,10){\makebox(0,0)[l]{$-x$}}
\end{picture}
\end{center}
\caption{Modified quantum circuit for the implementation of the Deutsch
algorithm on an NMR quantum computer.  Hadamard gates have been replaced by
pseudo-Hadamard and inverse pseudo-Hadamard gates, that is $90^\circ_{\pm y}$
pulses. The final $90^\circ_y$ excitation pulses cancel out the
$90^\circ_{-y}$ pulses, and thus all four pulses can be omitted.}
\label{fig:fdnmr}
\end{figure}
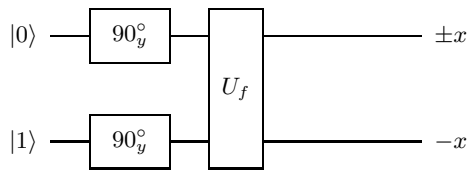

Both classical function analysis and Deutsch's algorithm were implemented,
using the modified quantum circuits shown in figures \ref{fig:f01nmr} and
\ref{fig:fdnmr}.  In these figures the $f$-\CNOT\ gates have been written as
general two qubit propagators, $U_f$, Hadamard gates have been replaced by
$90^\circ_{\pm y}$ pulses, and final $90^\circ_y$ pulses have been added at
the end of each sequence to convert eigenstates to observable magnetisation;
for simplicity pulses which simply act to cancel one another are omitted. The
initial pseudo-pure states were prepared using field gradient techniques
(spatial averaging), and the final results were determined by examining the
phase ($\pm x$) of the NMR signals from the two spins.

The heart of these computations is contained in the implementation of the
propagators $U_f$.  Each propagator corresponds to flipping the state of the
second spin as follows: $U_{00}$, never flip the second spin; $U_{01}$, flip
the second spin when the first spin is in state \ket{1}; $U_{10}$, flip the
second spin when the first spin is in state \ket{0}; $U_{11}$, always flip the
second spin. The first and last cases are particularly simple, as $U_{00}$
corresponds to doing nothing, while $U_{11}$ is just a selective $180^\circ_x$
pulse (a \NOT\ gate) on the second spin.  The second propagator is a \CNOT\
gate, while the third case is a reverse \CNOT\ gate, so that a \NOT\ gate is
applied to the target when the control spin is in state \ket{0}. These last
two gates can be implemented as described above.  The sequences actually used
were
\begin{equation}
90S_y - \frac{1}{4J} - 180_x - \frac{1}{4J} - 180_x 90I_y 90I_x 90_{-y}
90S_{\pm x}
\end{equation}
where pulses not marked as either $I$ or $S$ were applied to both nuclei. The
phase of the final pulse distinguishes $U_{01}$ (for which the final pulse was
$S_{+x}$) from $U_{10}$ (for which it was $S_{-x}$).  In retrospect it is
clear that these sequences are unnecessarily complicated; substantial
simplifications could have been achieved by combing pulses and by absorbing
phase shifts into abstract reference frames.

The second implementation of Deutsch's algorithm used the heteronuclear two
spin system provided by \nuc{13}{C}-labeled chloroform in solution in
deuterated acetone.  The initial pseudo-pure state was prepared by temporal
averaging, although results were also shown for computations beginning in the
thermal equilibrium state, while the results of the computation were
determined both by direct observation of the \nuc{1}{H} spectrum, and by full
quantum state tomography, which allows errors to be studied. Only the Deutsch
algorithm was demonstrated, with no results shown for classical computations.

The pulse sequences used to implement the $U_{01}$ and $U_{10}$ were similar
to those used for cytosine; slight differences can be traced to the
heteronuclear nature of the spin system and the consequent ability to place
\emph{both} spins exactly on resonance.  The sequences used to implement the
two balanced functions, $U_{00}$ and $U_{11}$ were more complicated than for
cytosine, as periods of free precession were included so that each of the four
propagators was applied over approximately the same time period (about
$1/2J$).  This means that the effects of relaxation (dominated by the
relatively short $T_2$ of the \nuc{13}{C} nucleus) were similar in all cases.

\subsection{Extensions and simplifications}
In addition to these two early examples several more implementations of
Deutsch's algorithm, and its more general cousin the Deutsch--Jozsa algorithm
\cite{Deutsch:1992}, have been published.  From among these a few particularly
interesting examples will be described in more detail.

The Deutsch--Jozsa algorithm is a generalisation of Deutsch's algorithm which
considers binary functions with any number of input bits.  Clearly such
functions need not be either constant (that is, give the same output for all
input values) or balanced (output 0 for half the possible inputs and 1 for the
remainder); for example a binary function with two input bits could return 0
for one of the four input values and 1 for the other three. Suppose, however,
that it is guaranteed that some (otherwise unknown) function $f$ is either
constant or balanced (such theoretically convenient if apparently arbitrary
guarantees are usually referred to as \emph{promises}), and it is necessary to
determine to which of these two categories the function belongs. To solve this
problem by classical means would in the worst case require the evaluation of
$f$ for just over half its inputs (if the function uses $n$ input bits it may
be necessary to evaluate the function over $2^{n-1}+1$ inputs), while even in
the best case it would be necessary to evaluate the function at least twice.
By contrast the Deutsch--Jozsa algorithm can always distinguish between
constant and balanced functions with a single function evaluation.

The first experimental implementation of the Deutsch--Jozsa algorithm
\cite{Linden:1998a}, which used functions with two input bits, is also notable
as the first example of the use of transition selective pulses.  The three
qubit system chosen was the homonuclear spin system made up by the three
\nuc{1}{H} nuclei in 2,3-dibromopropanoic acid.  For simplicity pseudo-pure
states were not prepared; instead the algorithm was simply applied to the spin
system in its thermal state.  Function evaluation for the six possible
balanced functions was accomplished by the simultaneous application of single
transition selective $180^\circ$ pulses to two of the four components of the
low field multiplet (which exhibits the largest separations between the four
components), and the result of the computation was determined by observation
of NMR signal intensities in the two high field multiplets.  This simple
approach worked remarkably well; the deviations from ideal behaviour seen in
experimental spectra were largely ascribed to the effects of strong coupling.

The Deutsch--Jozsa algorithm has also been applied to larger spin systems,
most notably a largely heteronuclear 5 spin system (containing single
\nuc{1}{H}, \nuc{15}{N} and \nuc{19}{F} nuclei, together with two \nuc{13}{C}
nuclei) derived from glycine \cite{Marx:2000}; in this case multiple pulse
techniques were used to implement controlled gates.  This system permits
functions with 4 input bits to be studied, but only one constant and one
(particularly simple) balanced function were actually implemented, out of a
possible total of two constant and 12870 balanced functions. Similarly while
an approach related to temporal averaging was used for initialisation, it was
used to construct an initial state which gave the same signal as a pseudo-pure
state, rather than the pseudo-pure state itself.  Thus this can only be
considered as a partial implementation.

Another important variation on the Deutsch--Jozsa algorithm is to remove the
ancilla qubit which is normally used to implement the phase trick, converting
function values into phase shifts.  This indirect approach is not necessary
for NMR implementations, as it is possible to implement controlled phase-shift
gates directly. Indeed it is usually \emph{simpler} to do this; in particular
the propagator for both constant functions is reduced to ``do nothing''. This
approach, sometimes called the refined Deutsch--Jozsa algorithm
\cite{Collins:1998}, has the advantage that one fewer qubit is required to
implement a given algorithm. It does however have one minor disadvantage for
demonstration algorithms, as such systems are intrinsically quantum mechanical
and \emph{cannot} be used to implement classical function analysis.

This simplified approach has been used with a three spin system (the three
\nuc{13}{C} nuclei in labeled alanine) to implement the Deutsch--Jozsa
algorithm for functions with three bit inputs \cite{Collins:1999}.  In this
case there are 70 balanced functions, from which ten representative functions
were chosen. At the other extreme this technique can also be used to implement
the refined Deutsch algorithm using a single qubit!  In this case the
algorithm is so simple as to be almost trivial: the $U_f$ propagator is a
$180^\circ_z$ pulse for the two balanced functions, while for the two constant
functions it is as usual ``do nothing''.  Since the $z$-pulse can be absorbed
into the RF reference frame, the pulse sequence can in principle be reduced to
a $90^\circ_y$ pulse followed by observation along $\pm x$.  Such a simple
experiment seems hardly worth performing, but for completeness it has been
demonstrated as one member of a set of experiments using the isolated one spin
and two spin \nuc{1}{H} systems is 5-nitro-2-furaldehyde \cite{Arvind:1999}.

\subsection{Grover's quantum search}
Grover's algorithm \cite{Grover:1997, Grover:1998} is designed to speed up
searches comparable to searching for a needle in a haystack.  More
mathematically it concerns the analysis of binary functions which map a large
number of bits to a single output bit, where the task is to determine an input
for which the value of the function is 1.  If the function has many inputs for
which its value is 1 (that is, if the haystack contains a substantial number
of needles), one of these inputs can be easily located by trial and error, but
if there is only one suitable input among a large number of unsuitable inputs
(one needle in a large haystack), locating this single input is obviously a
difficult process.

Suppose that the function $f$ has inputs described by $n$ bits, so that there
are $N=2^n$ possible inputs, and that $f=1$ for only one of these inputs.  The
only general way to locate this input is to evaluate $f$ over some trial
inputs, and look for a value $f=1$ (a satisfying input). A lucky guess would
permit this input to be located in one try, but this can hardly be relied on;
on average a random search (the best classical algorithm) would require about
$N/2$ trial evaluations, or $N-1$ evaluations in the worst case. The situation
is similar if there are $k$ inputs which satisfy the function; in this case
about $N/k$ trials will be required.  By contrast, Grover's algorithm permits
a satisfying value to be located with only $O(\sqrt{N/k})$ evaluations.  This
increase in computational efficiency is less impressive than the exponential
increase seen for Shor's quantum factoring algorithm \cite{Shor:1999}, but is
still quite substantial; furthermore the algorithm is quite general, with a
range of potential applications.

Early NMR implementations \cite{Chuang:1998b, Jones:1998b, Jones:1998c}
concentrated on the case $n=2$, so that there are 4 possible inputs, with only
a single satisfying input.  For this case the operation of the algorithm is
fairly simple to explain, and more complicated cases can be understood by
analogy. The algorithm involves two steps: evaluation of the function over all
possible inputs, followed by a selection process to pick out the desired
result.  As an example I will assume that $f(01)=1$ while $f(x)=0$ for all
other inputs $x$.

The algorithm begins with one quantum register (that is a group of qubits) in
a uniform superposition of the four possible inputs, so that its state is
\begin{equation}
\left(\ket{00}+\ket{01}+\ket{10}+\ket{11}\right)/2.
\end{equation}
An attempt to read out the value of this register will return one of the four
possible inputs at random.  A propagator implementing the function $f$ is then
applied, so that $f$ is evaluated over all 4 inputs; the propagator is set
(either directly or indirectly by means of the phase trick) to return the
value of $f$ as a phase shift, that is
\begin{equation}
\ket{x}\stackrel{U_f}\longrightarrow(-1)^{f(x)}\ket{x},
\end{equation}
so that at the end of the calculation the quantum register is in the state
\begin{equation}
\left(\ket{00}-\ket{01}+\ket{10}+\ket{11}\right)/2.
\end{equation}
The desired satisfying input has now in some sense been identified, as it
bears the unique mark of a negative phase; this is not, however, of any
immediate use, as an attempt to analyse this state will still return one of
the four contributing inputs at random.  It is, therefore, necessary to apply
some propagator which converts the \emph{phase} difference into an
\emph{amplitude} difference.

\begin{figure}
\begin{center}
\begin{picture}(235,80)
\put(10,60){\makebox(0,0)[r]{\ket{0}}} \put(15,60){\line(1,0){10}}
\put(25,50){\framebox(20,20){H}} \put(45,60){\line(1,0){15}}
\put(10,20){\makebox(0,0)[r]{\ket{0}}} \put(15,20){\line(1,0){10}}
\put(25,10){\framebox(20,20){H}} \put(45,20){\line(1,0){15}}
\put(60,10){\framebox(30,60){$U_{f_{ab}}$}} \put(90,60){\line(1,0){15}}
\put(105,50){\framebox(20,20){H}} \put(125,60){\line(1,0){15}}
\put(90,20){\line(1,0){15}} \put(105,10){\framebox(20,20){H}}
\put(125,20){\line(1,0){15}} \put(140,10){\framebox(30,60){$U_{f_{00}}$}}
\put(170,60){\line(1,0){15}} \put(185,50){\framebox(20,20){H}}
\put(205,60){\line(1,0){10}} \put(220,60){\makebox(0,0)[l]{\ket{a}}}
\put(170,20){\line(1,0){15}} \put(185,10){\framebox(20,20){H}}
\put(205,20){\line(1,0){10}} \put(220,20){\makebox(0,0)[l]{\ket{b}}}
\end{picture}
\end{center}
\caption{A quantum circuit for the implementation of Grover's quantum search
algorithm on a two qubit computer.  Boxes marked H are Hadamard gates.  The
first two qubit gate $U_{f_{ab}}$ corresponds to evaluation of the function
$f_{ab}$, replacing an eigenstate \ket{ij} by $-\ket{ij}$ if $i=a$ and $j=b$,
while $U_{f_{00}}$ simply replaces \ket{00} by $-\ket{00}$. \label{fig:cg}}
\end{figure}
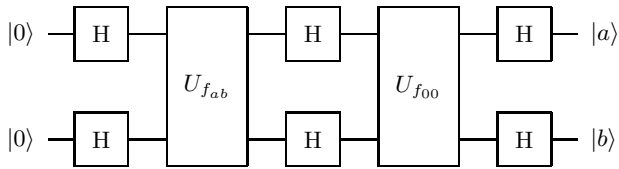

This process might seem simple, but it is in fact quite tricky, as any such
propagator must correspond to a logically reversible unitary operation.  There
is, however, a solution: inversion around the average.  This slightly peculiar
operation takes in a superposition and reflects the amplitude of each
component around the \emph{average} amplitude of all the components.  In the
example, the individual amplitudes are $\pm\half$, and the average amplitude
of the four components is $\frac{1}{4}$; reflecting $\half$ around
$\frac{1}{4}$ gives 0, while reflecting $-\half$ gives 1.  Thus this operation
acts to concentrate \emph{all} the amplitude on one member of the
superposition, giving a final state of just \ket{01}.  Surprisingly this
apparently complex operation can be performed using only Hadamard gates and a
controlled phase-shift gate which negates \ket{00} while leaving all other
states alone.  A quantum network for implementing this algorithm is shown in
figure \ref{fig:cg}; this network assumes that the function is evaluated using
propagators which apply the necessary phase shifts \emph{directly}; this can
of course be achieved using an ancilla qubit if desired, but direct
application of phase shifts has been the universal practice in NMR
implementations of this algorithm.

\subsection{NMR implementations}
The first implementation of Grover's search \cite{Chuang:1998b} was performed
using a heteronuclear NMR quantum computer based on chloroform.  This used
temporal averaging to prepare the initial pseudo-pure state, and quantum state
tomography to characterise the final result.  Quantum logic gates were
implemented using multiple pulse sequences as discussed below.

As this implementation used a heteronuclear spin system, both nuclei were
placed on resonance in their respective rotating frames; thus it was not
necessary to refocus chemical shifts, and periods of free precession
correspond to evolution under the spin--spin coupling.  The four desired
controlled phase shift gates were achieved (up to an irrelevant global phase)
by combining this with $\pm z$ rotations on the two spins; these were
explicitly implemented using composite $z$-pulses.  The Hadamard gates were
implemented using the two pulse sequence $90_{-y}180_x$ described in section
\ref{sec:NMRgates}.  Finally the elements of the pulse sequence with the
exception of the initial pair of Hadamard gates were assembled together to
give a single propagator, $U_{ab}$, and sequential pairs of pulses were
combined where ever possible to give simpler pulse sequences.  Thus the final
pulse sequences for $U_{ab}$ were
\begin{equation}
\frac{1}{2J} - 90_{-y} 90S_{\pm x} 90I_{\pm x} - \frac{1}{2J} - 90_{-y}
90_{-x}
\end{equation}
where, as before, pulses not marked as either $I$ or $S$ were applied to both
nuclei.  The choice of $\pm$ signs on the second and third pulses determined
which of the three functions $f$ is implemented.  Results were shown only for
$U_{11}$, in which case both signs are positive, but experiments were
performed for all four functions.

The second NMR implementation of Grover's quantum search \cite{Jones:1998b}
was based on the homonuclear \nuc{1}{H} spin system in cytosine.  In this case
spatial averaging was used to prepare the initial pseudo-pure state, and the
result was analysed by excitation and detection of the two \nuc{1}{H} signals;
this provides a particularly simple and immediate readout scheme.  Quantum
logic gates were implemented using multiple pulse sequences, with soft pulses
used to perform selective excitation and spin echoes used to refocus chemical
shifts; sequential pairs of pulses were combined within the individual
propagators $U_{f_{ab}}$, but no attempt was made at global simplification of
the pulse sequence.  For these reasons the homonuclear implementation produced
relatively poor results, although the cosmetic appearance of the spectra was
greatly improved by the application of a magnetic field crush gradient between
the end of the quantum circuit and the application of a final $90^\circ$ pulse
prior to detection.

\subsection{Extensions}
Grover's quantum search algorithm is not, of course, limited to two qubit
implementations, but can be used to search over a space described by any
number of input qubits, $n$, in which case there are $N=2^n$ possible inputs.
For $n>2$ the behaviour of the algorithm is similar to but more complex than
that described above; a single application of $U_{ab}$ (that is, function
evaluation and inversion around the average) acts to concentrate the intensity
of the superposition on the satisfying state, but does not simply produce this
state. Instead it is necessary to apply these two operations repeatedly,
driving the register towards the desired state.  The intensity of the desired
state oscillates with a frequency inversely proportional to $\sqrt{N}$, and so
after $O(\sqrt{N})$ applications of $U_{ab}$ the intensity will be largely on
the satisfying state; measurement of the register will then return this state
with high probability. Further application of $U_{ab}$ will then drive the
register \emph{away} from the desired state, and so it is important to choose
the correct number of repetitions.  This oscillatory behaviour was in fact
demonstrated for the two qubit case in the first NMR implementation
\cite{Chuang:1998b}, where $U_{ab}$ was applied between zero and seven times.
More recently Grover's algorithm has been implemented with three qubits,
searching over eight possible inputs, with up to 28 repetitions of the
propagator \cite{Vandersypen:2000}; this implementation used the
\nuc{1}{H}--\nuc{19}{F}--\nuc{13}{C} spin system in \nuc{13}{C} labeled $\rm
CHFBr_2$.

Another variant on Grover's algorithm occurs when there is more than one
satisfying input; in this case the number of such inputs is usually called
$k$.  The algorithm is almost identical to the simple case when $k=1$, except
that it is only necessary of use $O(\sqrt{N/k)}$ repetitions to drive the
quantum register into an equally weighted superposition of the $k$ satisfying
inputs. A measurement on this register will then cause the superposition to
collapse into one of its constituent values, and so one of the $k$ satisfying
inputs can be returned at random.  Clearly this requires either that the value
of $k$ be known beforehand, or that it be determined; fortunately $k$ can be
readily estimated using an extension of Grover's search called quantum
counting \cite{Jones:1999a, Boyer:1996, Brassard:1998, Mosca:1998}.

The basic idea behind quantum counting is that in addition to driving the
quantum register towards the satisfying values application of the Grover
propagators also results in a phase shift which depends on the value of $k/N$.
In a conventional implementation of Grover's algorithm this phase shift is a
\emph{global} phase shift, and so cannot be detected.  Quantum counting,
however, uses an approach similar to Deutsch's algorithm to convert this phase
shift into a \emph{relative} phase shift which can be measured.  This
algorithm requires one additional qubit, and so when it was implemented on the
cytosine system \cite{Jones:1999a} it was performed using one bit functions,
for which either zero, one, or two inputs satisfy the function.

\section{Quantum phenomena} \label{sec:phenomena}
In addition to implementing quantum computations, NMR techniques have also
been used to implement other more general aspects of quantum information
processing, including demonstrations of some quantum phenomena.  A few of the
more important examples are described below.

\subsection{Entangled states}
Entangled states \cite{Bouwmeester:2000b} are states of quantum mechanical
systems which cannot adequately be characterised by describing the states of
their component subsystems; instead the properties of such states are
properties of the system as a whole.  An important example is provided by the
four Bell states:
\begin{equation}
\phi^\pm=\frac{\ket{00}\pm\ket{11}}{\sqrt{2}}\qquad
\psi^\pm=\frac{\ket{10}\pm\ket{01}}{\sqrt{2}}.
\end{equation}
(The cat states introduced in Section \ref{sec:NMRinit} are an obvious
generalisation of $\phi^\pm$ to systems of more than two qubits.)  Such states
play a central role in experimental tests of quantum theory, and also form the
essential information processing resource for quantum communication
techniques.  For this reason there is significant interest in techniques for
generating entangled systems, especially multiple particle entangled systems
such as cat states.

The quantum network for generating cat states is well known, and the three
qubit version is shown in figure \ref{fig:cat}.  This was used early on to
generate Bell states and three qubit cat states (GHZ states)
\cite{Laflamme:1998}, and more recently has been used to prepare seven qubit
cat states \cite{Knill:2000}.  There are, however, several reasons for
questioning the true significance of these results, and they have certainly
not generated as much excitement as similar results with smaller numbers of
qubits in other quantum technologies \cite{Bouwmeester:1999}.

The first reason for skepticism is the close relationship between cat states
and multiple quantum coherences.  As described above (Section
\ref{sec:NMRinit}) $nQ_x$ multiple quantum coherence corresponds to a mixture
of \ket{\phi_n^+}\bra{\phi_n^+} and \ket{\phi_n^-}\bra{\phi_n^-}; thus the
generation of high order multiple quantum coherence is nearly equivalent to
the production of cat states.  Seen in this light, seven quantum coherence is
not particularly impressive: solid state NMR techniques have been used to
generate coherence orders above one hundred \cite{Emsley:1992}.

A second difficulty with NMR cat states arises from the use of pseudo-pure
states.  As discussed in Section \ref{sec:NMRqm}, the fact that NMR density
matrices are always highly mixed, with nearly equal populations in all spin
states, appears to mean that they cannot, strictly speaking, exhibit
entanglement. Thus NMR cat states might be more properly described as
\emph{pseudo-entangled} states.

Finally even if truly entangled NMR states were to be produced there are are
serious limitations on the use of NMR to investigate quantum mechanics.  The
ensemble nature of NMR measurements complicates the investigation of
deviations from classical behaviour, while the short distances over which NMR
entanglement can be produced (normally confined to molecular dimensions)
compares unfavourably with the distance achievable with entangled photons
(hundreds of metres).

\subsection{Quantum teleportation}
Quantum teleportation \cite{Bouwmeester:2000c, Bennett:1993, Bouwmeester:1997}
is a particularly intriguing example of quantum communication; in essence it
involves the transfer of an unknown quantum state from one quantum particle to
another, without any attempt to characterise the state.  The technique relies
on the peculiar, apparently non-local, correlations inherent in entangled
states, such as Bell states. Note that quantum teleportation does not permit
the direct transfer of a quantum system into empty space: a suitable target
particle (one half of a Bell state) must be provided at the destination to
receive the quantum information.  Thus it is the \emph{state} of the particle
which is teleported, and not the particle itself.

Quantum teleportation has been implemented on a three spin NMR quantum
computer \cite{Nielsen:1998} using two \nuc{13}{C} nuclei and a single
\nuc{1}{H} nucleus in \nuc{13}{C} labeled tri\-chloro\-ethene.  The process
can be summarised by
\begin{equation}
\ket{\psi_a}\left(\ket{0_b 0_c}+\ket{1_b 1_c}\right) \longrightarrow
\left(\ket{0_a 0_b}+\ket{1_a 1_b}\right)\ket{\psi_c}
\end{equation}
where \ket{\psi} indicates an arbitrary quantum state, and the subscripts $a$,
$b$, $c$ simply label the three qubits.  Although this implementation raises
some interesting issues, arguments similar to those used above for entangled
states can be applied to NMR teleportation, and once again optical
implementations \cite{Bouwmeester:1997} are rather more convincing.

\subsection{Error correction}
Any computing technology is ultimately based on some physical device, and such
devices are inevitably error prone.  There are two main methods by which the
effects of these random errors can be reduced: stabilization techniques which
act to cancel out the effects of small errors, and error correction techniques
which detect, characterise and finally fix the results of larger errors.

Stabilization against small errors is an inherent feature of digital
information processing.  In any digital system information is stored as ones
and zeroes, which are ultimately represented as two states of a physical
system, such as high and low voltages.  Small fluctuations away from the two
ideal voltages (noise) do not matter, as long as it is always clear whether
the voltage is high or low.  In some cases this passive insensitivity to noise
can be further enhanced by active stabilization, which acts to continuously
drive the system towards the nearer of the two ideal states.  Because of this
intrinsic stabilization digital information processing devices are effectively
invulnerable to the effects of noise, as long as the noise signals remain
below some critical threshold.  If the noise rises above this threshold,
however, stabilization is no longer effective, and it is necessary to resort
to error correction.

Error correction techniques are even more important for quantum information
processing, as simple stabilization techniques are ruled out.  Unlike bits
qubits are not confined to two states, but can also exist in superpositions of
these states, and any stabilization scheme which drives a qubit back towards
the two eigenstates will of course destroy these vital superpositions.  For
some time it was believed that the nature of superposition states would also
render error correction schemes impractical, but happily this is not in fact
the case.

Classical error correction schemes \cite{Feynman:1996} are most simply
described in terms of the transmission of information along a noisy channel,
which has the effect of flipping bits (that is, changing them from 0 to 1 and
\emph{vice versa}) at random, with an error probability $\epsilon$.  Quantum
error correction schemes are similar except that as well as correcting qubit
flip errors (that is errors which take a qubit from \ket{0} to \ket{1}) it is
also necessary to correct phase errors (which interconvert, for example,
$\ket{0}+\ket{1}$ and $\ket{0}-\ket{1}$).  For simplicity, however, I will
only describe how classical schemes work.  All such schemes (including quantum
schemes \cite{Shor:1995, Steane:1996, Steane:1998, Steane:1999}) use multiple
copies of each bit (or qubit); the redundancy provided by these ancillas
allows errors to be detected and corrected.

The simplest classical scheme is triplet coding, in which a bit is encoded
using three repetitions (so that 0 is encoded as 000, while 1 is encoded as
111) and decoded by taking a majority vote (so that 000, 001, 010 and 100 all
decode as 0, while 111, 110, 101 and 011 all decode as 1). This scheme is
robust against random errors in any \emph{one} of the three bits.  Of course
if there are errors in two of the bits then the message will still be
corrupted; however the chance of two errors occurring is $\epsilon^2
(1-\epsilon)$, and as long as $\epsilon$ is small this possibility can be
neglected.  If the level of errors is too high then triplet coding is no
longer effective, but more robust schemes (involving even greater redundancy)
can be used.

\begin{figure}
\epsfig{file=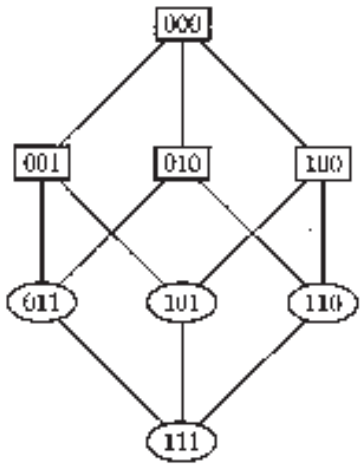} \caption{The relationship between the triplet error
correction code and active stabilization.  The eight possible states of three
bits can be placed on the eight corners of a cube, where the sides of the cube
connect states which differ by a single bit flip and the two ideal states (000
and 111) lie at opposite corners. Decoding a triplet state by majority vote is
equivalent to moving the state to the nearer of the two ideal corners.}
\label{fig:triplet}
\end{figure}

Triplet coding and other error correction schemes might seem very different
from stabilization schemes, but in fact the basic ideas are quite similar as
shown in figure \ref{fig:triplet}.  In effect the code divides the eight
possible settings of a three bit system into two subspaces, just as voltages
can be divided into high and low.  Note that if a simpler doublet coding
scheme (in which 0 is represented as 00 and 1 as 11) is used it is still
possible to \emph{detect} single bit errors, but not to \emph{correct} them,
as the two states which can occur after a bit flip error (01 and 10) lie
equally close to the two ideal states.  In communication (as opposed to data
storage) schemes, however, it may be sufficient to detect errors, as the
erroneous bits can then be sent again.

Some simple quantum error detection and correction protocols have been
implemented on NMR quantum computers \cite{Cory:1998b, Leung:1999}.  Full
quantum error correction requires at least five qubits to encode a single
state, but simpler schemes exist which use only three qubits; these simplified
schemes can only correct phase errors or bit-flip errors, but not both.  Phase
errors occur as a result of spin--spin relaxation, while bit-flip errors
correspond to spin--lattice relaxation, and so in many NMR systems phase
errors will dominate.  Furthermore, phase errors only occur in quantum
computers, as they have no classical analogue.  For these reasons early
studies on NMR error correction have concentrated on three qubit
phase-correcting codes.

It should be noted that these NMR experiments are demonstrations of the
principle of error correction, rather than practical implementations of error
correcting codes.  In order to effectively suppress errors in a quantum
computation it is necessary to apply the error correction protocol repeatedly;
this in turn requires that the ancilla qubits be maintained in the correct
state.  This is most simply achieved by initialising them to \ket{0} before
each correction round.  Unfortunately the NMR techniques described in Section
\ref{sec:NMRinit} allow qubits to be initialised only once, at the start of
the calculation; they cannot be repeatedly reinitialised.  This appears to
rule out current NMR implementations as practical technologies for quantum
computation \cite{Jones:2000d}.

\section{NMR and entanglement}\label{sec:NMRqm}
Finally I will return to the question, briefly discussed in Section
\ref{sec:building}, of whether NMR quantum computers are in fact true quantum
computers at all.  Much of the opposition to NMR as a quantum computing
technology stems from the formidable difficulties \cite{Jones:2000d} in
scaling up the current small systems to computers with a reasonable number of
qubits. A particularly common observation is that the use of pseudo-pure
states is exponentially inefficient \cite{Warren:1997, Gershenfeld:1997b}, in
that the amount of pseudo-pure state which can be extracted from an NMR system
at thermal equilibrium falls off exponentially with the number of spins in the
system. Of course this problem can be overcome by using an exponentially large
sample, but this approach would remove any increase in computational
efficiency supposedly arising from quantum mechanical effects: there are a
wide range of classical techniques (such as DNA computing \cite{Adleman:1994})
which allow exponential gains in computing power to be obtained from
exponentially large samples.

This problem is not in principle unique to NMR; it will occur in any potential
quantum computing technology which works in the high temperature limit
\cite{Jones:2000d}.  It can in principle be overcome by working at
sufficiently low temperatures, or by using some other initialisation technique
to produce a non-Boltzmann population distribution, although the technical
problems involved are substantial \cite{Jones:2000d}. However NMR \emph{is}
the only technology among those currently under investigation which falls into
this category.

In addition to the obvious technological issues raised by this exponential
efficiency there are also some more fundamental concerns.  It has long been
suspected that the non-classical power of quantum computation is closely
linked to the existence of entangled states during quantum computations
\cite{Ekert:1998}.  Although this belief has never actually been proved, and
some recent theoretical results have suggested that it may not be entirely
correct \cite{Knill:1998b}, it is clear that some important algorithms such as
Shor's quantum factoring algorithm do require the generation of entangled
states \cite{Linden:1999d}. It can be shown that the pseudo-entangled states
generated in NMR quantum computations do not actually fulfill the mathematical
requirements for true entanglement \cite{Braunstein:1999}, casting doubt on
their ability to exhibit true quantum phenomena.  As we shall see, however,
this concern may not be entirely well founded.

\subsection{Quantifying entanglement}
Although the concept of entanglement is relatively easy to explain, actually
quantifying the amount of entanglement in any given system is a surprisingly
difficult task.  For a system of two qubits in a pure state the problem is
relatively straightforward, and the four Bell states ($\phi^\pm$ and
$\psi^\pm$, see Section \ref{sec:phenomena}) form a basis set describing the
possible \emph{maximally entangled} states.  With larger systems the problem
is much more difficult, as different definitions of entanglement lead to quite
different conclusions.

Difficulties can also arise when considering \emph{mixed states}, such as
those observed in NMR experiments.  These difficulties occur because there is
no unique way to break down a given mixed state into a mixture of pure states.
To see this consider the \emph{maximally mixed} state, which for a two qubit
system has the form
\begin{equation}
\frac{\bm{1}}{4}=
\begin{pmatrix} \quarter&0&0&0 \\ 0&\quarter&0&0 \\
                 0&0&\quarter&0 \\ 0&0&0&\quarter
\end{pmatrix}.
\end{equation}
This is perhaps most easily described as an equally populated mixture of the
four eigenstates, but this is by no means the only possible description: an
equally weighted mixture of the four Bell states will have exactly the same
form!  It might be claimed that this alternative decomposition is unnatural,
but there are no real grounds for such a statement, as the choice of basis set
is entirely arbitrary, and this approach is sometimes referred to as the
\emph{preferred ensemble fallacy}.

Similar difficulties arise when considering the pseudo-entangled states formed
from pseudo-pure states, such as
\begin{equation}\label{eq:psent}
(1-\epsilon)\frac{\bm{1}}{4}+\epsilon\ket{\psi^+}\bra{\psi^+} =\begin{pmatrix}
\frac{1+\epsilon}{4}&0&0&\frac{\epsilon}{2}\\ 0&\frac{1-\epsilon}{4}&0&0 \\
0&0&\frac{1-\epsilon}{4}&0 \\ \frac{\epsilon}{2}&0&0&\frac{1+\epsilon}{4}
\end{pmatrix}
\end{equation}
(mixed states of this kind were first considered by Werner \cite{Werner:1989},
and thus are often referred to as Werner states). It is tempting to argue that
this mixed state \emph{is} a mixture of the maximally mixed state together
with a fraction $\epsilon$ of an entangled state, but as discussed above there
is no particular reason to choose this description.  The amount of apparent
entanglement in the state could be increased by decomposing the maximally
mixed state as a mixture of entangled states, but it is also possible to
choose decompositions which \emph{reduce} the apparent contribution from
entanglement.

While it is not possible to say how much entanglement is in any particular
mixed state it is possible to determine the minimum contribution from
entangled states which must be present in the state (in a pure entangled state
this minimum fraction is, of course, one). If this minimum quantity is greater
than zero then it is reasonable to say that the mixed state does contain some
entanglement; if, however, the minimum amount is zero then it is possible to
describe the mixed state using only product states (and thus without invoking
entanglement), and the state is said to be \emph{separable}.  In particular it
can be shown \cite{Peres:1996} that pseudo-entangled states are in fact
separable if $\epsilon < 1/3$.  An explicit separable decomposition of
equation \ref{eq:psent} with $\epsilon=1/9$ is given in appendix
\ref{ap:psent}.

With larger systems the problem is more complicated, but two important results
are known \cite{Braunstein:1999}.  Given any state \ket{\psi} of an $n$-qubit
system, a mixture made by mixing a fraction $\epsilon$ of this state into the
maximally mixed state, $\bm{1}/2^n$, can always be shown to be separable for
sufficiently small values of $\epsilon$, such that
\begin{equation}\label{eq:lower}
\epsilon\le\frac{1}{1+2^{2n-1}}\sim\frac{2}{4^n}.
\end{equation}
It can also be shown that non-separable states \emph{do} exist for
sufficiently large values of $\epsilon$, such that
\begin{equation}\label{eq:upper}
\epsilon>\frac{1}{1+2^{n/2}}\sim\frac{1}{2^{n/2}}.
\end{equation}
By comparison the values of $\epsilon$ obtainable with NMR quantum computers
working within the high temperature limit \cite{Jones:2000d, Warren:1997,
Gershenfeld:1997b} are given by
\begin{equation}
\epsilon\sim\frac{n}{2^n}
\end{equation}
which lies between the two bounds given in equations \ref{eq:lower} and
\ref{eq:upper}.  Using realistic parameters it can be calculated that the
states used in NMR quantum computations are always separable for systems with
less than about $13$ qubits and may (or may not) become entangled beyond this
point.  Since the systems used so far have involved no more that seven qubits,
all NMR quantum computations to date have involved purely separable states.

\subsection{NMR and quantum mechanics}
The observation that NMR quantum computers have so far only used separable
states has led some authors to suggest that they are not true quantum
computers at all!  When assessing claims of this kind it must be remembered
that \emph{quantum} is being used here in its technical sense of
\emph{provably non-classical}.  Consider, for example, a set of NMR quantum
computers which are identical except for having different values of
$\epsilon$, with some of them lying above the entanglement limit discussed
previously, and the remainder lying below this limit.  It seems very strange
to claim that two groups of computers are fundamentally different in
character, with the first group being quantum mechanical and the second group
classical, but it is more reasonable to suggest that \emph{only} the computers
in the first group are capable of exhibiting \emph{convincingly non-classical}
behaviour.

Even this claim, however, may be too strong.  It seems highly unlikely that it
is the mere presence of entanglement which leads to non-classical efficiency;
rather it is the ability to interconvert a wide range of entangled and
non-entangled states.  Thus in order to claim that NMR quantum computing
experiments are classical it is not sufficient to show that they involve only
classical states; instead it is necessary to show that the processes which
connect these states can themselves be described classically.  To date
attempts to achieve this have failed \cite{Schack:1999}, and it is not clear
that such a model is possible.

In an unrelated approach, some authors have attempted to draw a distinction
between the density matrix (which is a description of the state of an NMR
system) and the state itself, although it is tricky to draw this distinction
in an entirely convincing fashion.  It is true, however, that the density
matrix approach is only an approximate description of an NMR system, and that
any conclusions based on this approximation are to some extent open to
suspicion.

\section{Conclusions}
When assessing NMR quantum computation it is important to take a balanced
view, avoiding both excessive excitement at the apparently impressive results
achieved so far and undue despair at the limitations that have been
identified.  NMR quantum computation has been the subject of a great deal of
skeptical scrutiny, probably more than any other approach. In part this is a
result of the great success of NMR as a technique for quantum information
processing; furthermore, the highly developed nature of NMR experiments, in
comparison with many other putative quantum technologies, means that the
limits of the technique are well known and understood.

On the positive side, NMR is far ahead of any competing technology in the
implementation of quantum computations and other forms of quantum information
processing.  Although some basic elements have been implemented using other
technologies, such as the ion trap \CNOT\ gate \cite{Monroe:1995}, NMR remains
the only technology capable of implementing a complete quantum algorithm.
Progress from two qubit devices \cite{Cory:1996, Cory:1997, Gershenfeld:1997,
Jones:1998a, Chuang:1998a} to systems with seven qubits \cite{Knill:2000} has
been extremely rapid, and there is every reason to believe that more progress
will soon be made.

Against this it must be pointed out that most researchers believe that the
current designs for NMR quantum computers cannot be extended very much
farther: while there is some disagreement as to which technical difficulty
will actually stop further progress it is widely agreed that it will be
difficult to progress beyond 10--20 qubits.  While current demonstration
systems are undoubtedly interesting they are far too small to be used to
tackle problems beyond the range of current classical computers.  Similarly,
although there are important applications of quantum information processing,
such as quantum cryptography \cite{Ekert:2000b}, which require only devices
with small numbers of qubits, all such applications lie in the field of
quantum communication where NMR methods appear completely unsuitable.

The discovery \cite{Braunstein:1999} that current NMR implementations of
quantum computation do not seem to involve entanglement might appear a serious
blow, but its implications should not be overstated.  This result does
\emph{not} mean that NMR quantum computers are not true quantum computers,
although it \emph{does} appear to mean that they cannot be used to achieve
non-classical efficiencies.  However it has long been known \cite{Warren:1997,
Gershenfeld:1997b} that the exponential inefficiency of pseudo-pure state
preparation means that current implementations are unlikely to exhibit true
efficiency gains.  It seems likely that in the next few years it will be
possible to use \emph{para}-hydrogen techniques \cite{Natterer:1997,
Duckett:1999, Hubbler:2000} to build two qubit NMR quantum computers above the
entanglement threshold, but it will be tricky to extend this approach to
larger systems.

\section*{Acknowledgements}
I thank Mark Bowdrey, Holly Cummins, Ruth Dixon and Vlatko Vedral for helpful
conversations, and Steffen Glaser for providing a preprint of his work
\cite{Hubbler:2000}.

\appendix

\section{An explicitly separable decomposition of a pseudo-entangled
state}\label{ap:psent}
While it can be shown that a two qubit pseudo-entangled
state with $\epsilon<1/3$ is in fact separable \cite{Peres:1996}, the argument
used to derive this result does not provide an explicit decomposition of such
states into product states, but merely proves that such a decomposition
exists.  It is, however, fairly simple to find such a decomposition
\cite{Braunstein:1999} for mixed states with low values of $\epsilon$.

The process begins by constructing an \emph{overcomplete basis} for a single
qubit; a basis set of this kind is sufficient to describe any state of a
single qubit (a single spin), but contains more basic elements than is
strictly necessary.  One suitable basis is the set of six states
\begin{equation}
\ket{0}\quad \ket{1}\quad \frac{\ket{0}\pm\ket{1}}{\sqrt{2}}\quad
\frac{\ket{0}\pm i\ket{1}}{\sqrt{2}}
\end{equation}
which may be labeled as $\beta_j$, $j=1,\,2,\,\dots 6$.  This can then be used
to construct a two qubit basis set, $\beta_{jk}$, by taking direct products.
This basis set, comprising 36 elements, was constructed by taking products of
single qubit states, and so is explicitly composed of product states
\emph{only}.  Thus any density matrix which can be decomposed in this basis
must be separable.

As an example consider a pseudo-entangled state of the form given by equation
\ref{eq:psent} with $\epsilon=1/9$:
\begin{equation}\label{eq:ninth}
\rho=\frac{2}{9}\bm{1}+\frac{1}{9}\ket{\psi^+}\bra{\psi^+}=
\frac{1}{18}\begin{pmatrix}5&0&0&1\\0&4&0&0\\0&0&4&0\\1&0&0&5\end{pmatrix}.
\end{equation}
This can be written in the form
\begin{equation}
\rho=\sum_{jk}P_{jk}\beta_{jk}
\end{equation}
and the proportions $P_{jk}$ of each basis state $\beta_{jk}$ can be obtained
from the trace of the product of $\rho$ and $\beta_{jk}$ giving
\begin{equation}
P=\frac{1}{36}\begin{pmatrix} 2&0&1&1&1&1 \\ 0&2&1&1&1&1 \\ 1&1&2&0&1&1 \\
1&1&0&2&1&1 \\ 1&1&1&1&0&2 \\ 1&1&1&1&2&0
\end{pmatrix}.
\end{equation}
This calculation also makes clear the danger of an unquestioning equation of
entanglement and multiple quantum coherence.  The state shown in equation
\ref{eq:ninth} clearly contains double quantum coherence (it can be decomposed
in product operator notation as $\half E/2+2I_zS_z/18+DQ_x/9$) and yet is not
provably entangled.

\end{document}